\documentclass[aps,prl,twocolumn,superscriptaddress,preprintnumbers]{revtex4-1}
\usepackage{epsf,epsfig}
\usepackage[utf8]{inputenc}
\usepackage{amssymb,amsmath,amsfonts}
\usepackage{color}
\usepackage{slashed}
\usepackage{tensor} 
\usepackage{braket}
\usepackage{float}
\usepackage[normalem]{ulem} 
\usepackage[colorlinks=true]{hyperref}  
\hypersetup{
    bookmarks=true,         
    unicode=false,          
    pdftoolbar=true,        
    pdfmenubar=true,        
    pdffitwindow=false,     
    pdfstartview={FitH},    
    colorlinks=true,       
    linkcolor=magenta, 
    citecolor=blue,        
    filecolor=magenta,      
    urlcolor=cyan           
} 

\newcommand{\bea}{\begin{eqnarray}}

\newcommand{\diff}{\mathrm{d}}
\newcommand{\eea}{\end{eqnarray}}
\usepackage{lipsum} 
\newcommand{\ba}{\begin{eqnarray}}
\newcommand{\ea}{\end{eqnarray}}

\newcommand{\beq}{\begin{equation}}
\newcommand{\eeq}{\end{equation}}
\newcommand{\beqa}{\begin{eqnarray}}
\newcommand{\eeqa}{\end{eqnarray}}
\newcommand{\beqar}{\begin{eqnarray*}}
\newcommand{\eeqar}{\end{eqnarray*}}

\usepackage{color}

\begin{document}

\title{Universal aspects of holographic quantum critical transport with self-duality}
\author{\'Angel J. Murcia}
\email{angel.murcia@pd.infn.it}
\affiliation{INFN, Sezione di Padova, Via Marzolo 8, 35131 Padova, Italy}

\author{Dmitri Sorokin} 
\email{dmitri.sorokin@pd.infn.it}
\affiliation{INFN, Sezione di Padova, Via Marzolo 8, 35131 Padova, Italy}
\affiliation{Dipartimento di Fisica ed Astronomia ``Galileo Galilei'',
Università degli Studi di Padova, Via Marzolo 8, 35131 Padova, Italy}


\begin{abstract}
We prove several universal properties of charge transport in generic CFTs holographic to nonminimal extensions of four-dimensional Einstein-Maxwell theory with exact electromagnetic duality invariance. First, we explicitly verify that the conductivity of these theories at zero momentum is a universal frequency-independent constant. Then, we derive their analytical expressions for non-zero momentum in any holographic duality-invariant theory for large frequencies and in the limit of small frequencies and momenta. Next, in the absence of terms that couple covariant derivatives of the curvature to gauge field strengths, two universal features are proven. On the one hand, it is shown that for a general-relativity neutral black-hole background the conductivities at any frequency and momentum are independent of the choice of duality-invariant theory, thus coinciding with those in the Einstein-Maxwell case. On the other hand, if higher-curvature terms affect the gravitational background, the conductivities get modified, but the contributions from nonminimal couplings of the gauge field to gravity  are subleading. We illustrate this feature with an example. 

\end{abstract} 

\maketitle

The AdS/CFT correspondence \cite{Maldacena:1997re,Gubser:1998bc,Witten:1998qj} has become a powerful and fruitful tool for the study of strongly-coupled systems in the vicinity of quantum critical points, leading to the development of the so-called AdS/Condensed Matter (AdS/CMT) duality \cite{Hartnoll:2009sz,Sachdev:2010ch,Zaanen:2015oix,Hartnoll:2016apf,Baggioli:2016rdj,nastase_2017}. Among other aspects, it has been possible to identify a variety of holographic models which exhibit properties characteristic of condensed-matter systems, such as superfluidity, superconductivity or (quantum) Hall conductivity \cite{Hartnoll:2008vx,Keski-Vakkuri:2008ffv,Herzog:2008he, Hartnoll:2008kx,Herzog:2009xv,Bergman:2010gm}. Further scrutiny of such interesting features have turned AdS/CMT into a highly active topic of research --- see \emph{e.g.} \cite{Hartnoll:2020fhc,Arean:2021tks,Ammon:2021pyz,Sword:2021pfm,Donos:2022www,Flory:2022uzp,Donos:2022qao}.

On the other hand, the potential of higher-order theories of gravity to unveil generic aspects of CFTs has become evident in the recent years. Apart from being able to capture finite $N$ and finite coupling effects within the canonical holographic correspondence between Type IIB String Theory and $\mathcal{N}=4$ Super-Yang-Mills theory \cite{Buchel:2004di,Benincasa:2005qc,Myers:2008yi}, higher-order gravities make it possible to explore holographic CFTs whose correlators take the most generic form allowed by conformal symmetry \cite{deBoer:2009pn,Buchel:2009sk,Myers:2010jv,Bueno:2018xqc,Cano:2022ord} or to identify new universal relations that hold for arbitrary CFTs \cite{Myers:2010xs,Perlmutter:2013gua,Mezei:2014zla,Bueno:2015rda,Chu:2016tps,Li:2018drw,Bueno:2018yzo,Bueno:2022jbl,Baiguera:2022sao}. These features have motivated the study of (charge) transport properties of holographic duals of higher-order gravities, observing novel and intriguing phenomena in the shear viscosity to entropy density ratio \cite{Kats:2007mq,Brigante:2007nu,Ge:2008ni}, holographic superconductivity \cite{Gregory:2009fj,Pan:2010at,Edelstein:2022xlb} and (electrical) conductivities \cite{Myers:2010pk,Pal:2010sx,Witczak-Krempa:2012qgh,Witczak-Krempa:2013aea,Ling:2016dck}.



In this work we explore various aspects of charge transport in CFTs holographic to
duality-invariant theories of electrodynamics with nonminimal couplings to gravity.  Duality invariance is a symmetry of the equations of motion of Einstein-Maxwell theory in vacuum, so it is justified to consider higher-order modifications that respect this symmetry. Explicit examples of duality-invariant theories are known to exist both with minimal couplings (see {\emph e.g.} \cite{Sorokin:2021tge} for a review) and with nonminimal couplings to gravity \cite{Cano:2021tfs,Cano:2021hje}.




In particular, we study CFTs holographic to duality-invariant theories whose equilibrium state is characterized by vanishing expectation values of all global charges (\emph{i.e.} systems without chemical potentials). We begin by explicitly checking that the conductivity at zero momentum is a frequency-independent universal constant. Then we are able to derive the explicit expressions for the conductivities in any holographic duality-invariant theory in the regimes of large frequencies and for sufficiently small frequencies and momenta. The latter depend on both the gravitational background selected and the duality-invariant theory under study, so that conductivities at non-zero momentum will generically differ with respect to their Einstein-Maxwell values.

Nonetheless, when dealing with CFTs holographic to duality-invariant theories which do not couple covariant derivatives of the curvature to gauge field strengths, it turns out that conductivities do possess two remarkable universal features. First, whenever a general-relativity black hole background is considered, the conductivities for any frequency and momentum are the same for all such holographic theories. Secondly, if the black hole background is modified by higher-curvature terms, the conductivities get corrected, but the contributions coming from nonminimal couplings between the curvature and the gauge field are subleading. We corroborate this feature with an explicit example.

\textbf{Duality-invariant bulk setup.} Let us consider generic nonminimal extensions of Einstein-Maxwell theory in four dimensions described by the following action:
\begin{equation}
 I=\kappa_N \int \diff^4 x \sqrt{\vert g \vert} \left[R+\frac{6}{L^2}-\chi^{\mu \nu \rho \sigma} F_{\mu \nu} F_{\rho\sigma}+\mathcal{L}_{\mathrm{grav}}^{\mathrm{high}}\right]\,,
    \label{eq:dualgen}
\end{equation}
where $\chi^{\mu \nu \rho\sigma}$ depends solely on the metric and the curvature, $\kappa_N=(16 \pi G)^{-1}$ and
\begin{equation}\label{Lhigher}
\mathcal{L}_{\mathrm{grav}}^{\mathrm{high}}=L^2 \sum_{i} \alpha_i^{(2)}\mathcal{R}_i^{(2)}+ L^4 \sum_{i}\alpha_i^{(3)} \mathcal{R}_i^{(3)}+\dots \,,
\end{equation}
where $\mathcal{R}_i^{(n)}$ stands for curvature invariants of $n$-th order --- the index $i$ denoting every such inequivalent term ---, $L$ is the cosmological-constant length scale and  $\alpha_i^{(n)}$ are dimensionless couplings characterizing the theory. Eq. \eqref{eq:dualgen} may also be interpreted as a generic effective action \cite{Weinberg:1995mt} obtained by adding to Einstein-Maxwell theory all possible terms quadratic in $F_{\mu \nu}$ which are compatible with diffeomorphism and gauge invariance. The reason why it suffices for our purposes to work at  $\mathcal{O}(F^2)$ will become apparent afterwards.



Let $\mathcal{T}_{\mu \nu}$ be a traceless and symmetric tensor constructed from  contractions of the curvature tensor and its covariant derivatives, and let $b_n$ be the coefficients appearing in the Taylor series $\sqrt{1+x^2}=1+\sum_{n=1}^\infty b_n x^{2n}$. If we take the tensor $\tensor{\chi}{_{\mu \nu}^{\rho \sigma}}$ in \eqref{eq:dualgen} to be\footnote{Observe that $\tensor{\chi}{_{\mu \nu}^{\rho \sigma}}$ can be compactly expressed as $\tensor{\chi}{_{\mu \nu}^{\rho \sigma}}=\tensor{\Theta}{_{\mu \nu}^{\rho \sigma}}+\sqrt{\tensor{\delta}{_{[\mu}^{[\rho}} \tensor{\delta}{_{\nu]}^{\sigma]}}+\tensor{\Theta}{^{2}_{\mu \nu}^{\rho \sigma}}}$, as in \cite{Cano:2021hje}.}
\begin{align}\nonumber
    \tensor{\chi}{_{\mu \nu}^{\rho \sigma}}&= \tensor{\delta}{_{[\mu}^{[\rho}} \tensor{\delta}{_{\nu]}^{\sigma]}}+\tensor{\Theta}{_{\mu \nu}^{\rho \sigma}}+\sum_{n=1}^\infty b_n \tensor{\Theta}{^{2n}_{\mu \nu}^{\rho \sigma}}\,,\\ \tensor{\Theta}{^{2n}_{\mu \nu}^{\rho \sigma}}&=\tensor{\Theta}{_{\mu \nu}^{\alpha_1 \alpha_2}}\tensor{\Theta}{_{\alpha_1 \alpha_2}^{\alpha_3 \alpha_4}} \cdots \tensor{\Theta}{_{\alpha_{4n-3} \alpha_{4n-2}}^{\rho \sigma}}\,, \label{chiTheta} \\ \nonumber \tensor{\Theta}{_{\mu \nu}^{\rho \sigma}}&=\tensor{\mathcal{T}}{_{[\mu}^{[\rho}} \tensor{\delta}{_{\nu]}^{\sigma]}}\,,
    \end{align}
then the action \eqref{eq:dualgen} describes the most general exactly duality-invariant theory of electrodynamics with nonminimal couplings to gravity and having at most quadratic terms in the Maxwell field strength \cite{Cano:2021hje}. More concretely, this means that the set of equations formed by the equations of motion of  \eqref{eq:dualgen} and by the Bianchi identity for $F_{\mu \nu}$ is invariant under rigid $\mathrm{U}(1)$ rotations of the complex tensor $F'_{\mu\nu}+iH'_{\mu\nu}=e^{i\alpha}(F_{\mu\nu}+iH_{\mu\nu})$, where
\begin{equation}
\star H_{\mu \nu}=\frac 12 \frac{\delta I}{\delta F^{\mu\nu}}=-\tensor{\chi}{_{\mu \nu}^{\rho \sigma}} F_{\rho \sigma}\,.
\end{equation}
 In the case of Maxwell theory minimally coupled to a higher-curvature gravity, $\tensor{\chi}{_{\mu \nu}^{\rho \sigma}}=\tensor{\delta}{_{[\mu}^{[\rho}} \tensor{\delta}{_{\nu]}^{\sigma]}}$ and $H_{\mu\nu}$ is just the Hodge dual of $F_{\mu\nu}$.

With the goal of studying thermal CFTs in flat Minkowski space, we consider gravitational backgrounds of  \eqref{eq:dualgen} (\emph{i.e.} we take $F_{\mu \nu}=0$) which correspond to AdS black holes with a planar horizon --- usually called  (AdS) black branes in the literature:
\begin{equation}
    \diff s^2=\frac{r_0^2}{\tilde{L}^2 u^2}\left (-N^2(u)f(u) \diff t^2+\diff x^2+\diff y^2 \right )+\frac{\tilde{L}^2}{u^2 f(u)} \diff u^2\,,
    \label{eq:bhbackground}
\end{equation}
where $r_0$ is a constant of dimension of length, $\tilde{L}$ denotes the AdS length scale, generically differing from the cosmological-constant scale $L$ because of the higher-order corrections \cite{Buchel:2009sk,Bueno:2016xff,Bueno:2016ypa}, and
\begin{equation}
   N=1+\tilde{N}\, , \quad  f=1-u^3+\tilde{f}\,,
\end{equation}
where $\tilde{N}$ and $\tilde{f}$ are $u$-dependent functions encoding the higher-order corrections with respect to the GR solution ($\tilde{N}=\tilde{f}=0$) such that $\lim_{u \rightarrow 0} \tilde{f}= \lim_{u \rightarrow 0} \tilde{N}=0$. In this coordinate system the AdS boundary is located at $u=0$, while the horizon (which we assume to exist) is at $u=u_h$. The black hole temperature is
\begin{equation}\label{BHT}
T=-\frac{r_0 f'(u_h)}{4 \pi \tilde{L}^2}\,.
\end{equation}

\textbf{Retarded correlators from the AdS/CFT correspondence.}
We are interested in computing the retarded two-point current correlator $C_{ab}$ of CFTs at finite temperature which are holographically dual to exactly duality-invariant theories quadratic in $F_{\mu \nu}$.


In a generic three-dimensional QFT with a current $J_a$ $(a=t,x,y)$, the retarded current-current correlator in momentum space $p^a=(\omega, \mathbf{k})$ is given by
\begin{equation}\label{Cab}
C_{ab}(p)=-i \int d^{3} x \, e^{-i p_a\, x^a} \Theta_H(t) \langle[J_a(x), J_b(0)] \rangle\,, 
\end{equation}
where $x^a=(t,x,y)$ are boundary coordinates and $\Theta_H(t)$ is the Heaviside step function.

Specifically, we will assume that the expectation values of all global conserved charges vanish in the equilibrium state, which is equivalent to exploring systems with no chemical potential \cite{Herzog:2007ij, Myers:2010pk}. In such a case, the correlator $C_{ab}$ can be derived holographically via studying linear perturbations $A_\mu$ which solve the classical equations of motion around a neutral black brane background \eqref{eq:bhbackground}. To this aim, we impose the gauge $A_u=0$ and decompose
the remaining non-vanishing components in momentum space:
\begin{equation}\label{FA}
\begin{split}
 A_a &(t,u,\mathbf{x})=\int  \frac{\diff^3 p}{(2 \pi)^3}  \, e^{-i \omega t +i \mathbf{k} \cdot \mathbf{x} }A_a (u,\omega,\mathbf{k})\,.
 \end{split}
\end{equation}
 Working in momentum space and taking the spatial momentum vector to be $\mathbf{k}=(k,0)$, the equations of motion for $A_a$ in a duality-invariant theory given by Eqs. \eqref{eq:dualgen} and \eqref{chiTheta}  can be expressed in the following compact form\footnote{Expressed in this way, it might appear that the equations of motion are of third order in derivatives. However, they can be seen to be equivalent to the (second-order) equations $\nabla_\mu \left ( \chi^{\mu \nu \rho\sigma} F_{\rho \sigma} \right)=0$ in the gauge $A_u=0$ by appropriate manipulations which allow to integrate them into second-order equations, uniquely fixed after requiring $A_\mu=0$ to be a solution.}:
\begin{align}
\label{eq:symA}
 \mathcal{S} \mathbf{A}'' -\mathcal{S}'\mathbf{A}'+ \frac{\tilde{L}^4}{r_0^2} \mathcal{S}^2\left ( \omega^2 \mathcal{S}-\frac{k^2}{\mathcal{B}}\right)\mathbf{A}&=0\,, \\
 \omega\, \mathcal{S}\, \mathcal{B} A_t' +k A_x'&=0\,,
 \label{eq:symAxAt}
\end{align} 
\if{}
\begin{align}
\label{eq:sym1}
q^2 \mathcal{C} A_t+ q  \mathfrak{w} \mathcal{C} A_x - (\mathcal{B} A_t')'   &=0\,, \\  
  \label{eq:sym2}   \mathcal{B} A_t' +\frac{q}{\mathfrak{w}\mathcal{C}} A_x'&=0\, , \\ \label{eq:sym3}
   q \mathfrak{w}\mathcal{C} A_t + \mathfrak{w}^2 \mathcal{C} A_x
+\left (\frac{A_x'}{\mathcal{C}}\right )'&=0\,, \\ \label{eq:sym4}
     A_y \left (  \mathfrak{w}^2 \mathcal{C}-\frac{q^2}{\mathcal{B}}\right )+ \left (\frac{A_y'}{\mathcal{C}}\right )'&=0\,,
\end{align} 
\fi
where prime denotes derivative with respect to $u$, $\mathbf{A}=(\mathcal{B}A_t',A_y)$ and where $\mathcal{B}$ and $\mathcal{S}$ are identified after evaluation of $\mathcal{T}_{\mu \nu}$ on the black brane background \eqref{eq:bhbackground} as follows:
\begin{align}
\label{eq:mostgendualinv}
  \hspace{-0.24cm}\left.  \mathcal{T}_{\mu}{}^{\nu}  \right \vert_{N,f}&=2(\theta+\varphi)\delta_\mu{}^t \delta_t{}^\nu+2(\theta-\varphi)\delta_\mu{}^u \delta_u{}^\nu-\theta\tensor{\delta}{_\mu^\nu}\,, \\ \theta&=\frac{N^2\mathcal{B}^2-1}{2N\mathcal{B}}\, , \quad \varphi=\frac{f^2 N^2 \mathcal{S}^2-1}{2fN \mathcal{S}}\,,
\end{align}
where we used that \eqref{eq:mostgendualinv} represents the most general form for a symmetric and traceless tensor built from the curvature of \eqref{eq:bhbackground} and its covariant derivatives (see appendix). The reason for the equations of motion of $A_\mu$ to take such a compact form is due to duality invariance \cite{Herzog:2007ij}.

Now, applying the holographic prescriptions originally presented in \cite{Son:2002sd,Policastro:2002se}, it is explained in the appendix that $C_{ab}$ can be obtained as follows:
\begin{equation}
    C_{ab}=-\left. \frac{4 r_0 \kappa_N}{ \tilde{L}^2}M'_{ab}\right \vert_{u=0}\,,
    \label{eq:corr0}
\end{equation}
where $M_{ab}$ is defined by\footnote{Boundary indices are lowered and raised with the Minkowski metric $\eta^{ab}$.} the relation $A_a=\tensor{M}{_{a}^{b}} A_b (0)$, with $A_b(0)=A_b \vert_{u=0}$. We impose infalling boundary conditions at the horizon for all components of $A_a$. This implies, on account of Eqs. \eqref{eq:symA} and \eqref{eq:symAxAt}, that 
both $\mathcal{B}A_t'$ and $A_x'/\mathcal{S}$ are proportional to $A_y$. Then, transforming these equations into an explicit second-order differential system for $A_a$ (see Footnote \cite{Note2}) and adapting the computations presented in \cite{Herzog:2007ij}, one may identify $M'_{ab}|_{u=0}$ and obtain the non-vanishing components of $C_{ab}$ from \eqref{eq:corr0}:
\begin{align}
\label{eq:clong}
    \frac{C_{tt}}{k^2}&=\frac{C_{xx}}{\omega^2}=-\frac{C_{tx}}{k \omega}=-\frac{C_{xt}}{k \omega}=\frac{4\tilde{L}^2 \kappa_N }{ r_0} \frac{A_y(0)}{A_y'(0)}\,,  \\
   C_{yy}&=-\frac{4 r_0 \kappa_N }{ \tilde{L}^2} \frac{A_y'(0)}{A_y(0)}\,. \label{eq:cyy}
\end{align}
 Having at our disposal the correlator $C_{ab}$, one may compute the so-called  longitudinal and transverse self-energies $K^L(\omega,\mathbf k)$ and $K^T(\omega,\mathbf k)$, defined as \cite{Herzog:2007ij}
\begin{equation}\label{KTL}
C_{xx}=-\frac{\omega^2}{\sqrt{k^2-\omega^2}} K^L\, ,\quad C_{yy}=\sqrt{k^2-\omega^2} K^T\,  .
\end{equation}
Comparing \eqref{eq:clong} with \eqref{eq:cyy}, we find that the product of $K^L$ and $K^T$ is the following universal constant for all frequencies and momenta:
\begin{equation}
K^L (\omega,\mathbf k) K^T(\omega,\mathbf k)=16 \kappa_N^2\,.
\label{eq:univresse}
\end{equation}
The above relation holds for all CFTs holographic to duality-invariant theories, since its derivation just requires to know their form up to quadratic order in the vector field, captured by the theories defined by Eqs. \eqref{eq:dualgen} and \eqref{chiTheta}. It matches\footnote{Up to a constant on the right-hand side of \eqref{eq:univresse}, which corresponds to the different conventions used for the electromagnetic duality transformation.} with the result obtained in \cite{Myers:2010pk} in the particular case of a background \eqref{eq:bhbackground} with $N=1$ after requiring duality invariance.





\textbf{Conductivity of holographic duality-invariant theories.} Following the usual holographic prescriptions, a gauge vector field on the bulk couples to a current on the boundary CFT. We are interested in studying the subsequent longitudinal and transverse conductivities $\sigma_x$ and $\sigma_y$ for the holographic theories defined by Eqs. \eqref{eq:dualgen} and \eqref{chiTheta}. According to the Kubo formula \cite{kubo1957statistical}, they are given by 
\begin{equation}
\label{sigmaTL}
\sigma_j (\omega,k)=-\mathrm{Im}\left ( \frac{C_{jj}}{\omega} \right)\,,\quad
j={x,y}\,.
\end{equation}
Particularly simple is the computation of  the conductivities at zero momentum $k=0$. In this case, spatial rotational invariance ensures that $K^L(\omega,0)=K^T(\omega,0)$ and $\sigma_x(\omega,0)=\sigma_y(\omega,0)$. Using \eqref{eq:univresse} and the expression for $C_{yy}$ given in  \eqref{eq:cyy}, one obtains\footnote{After choosing the sign for $K^L(\omega,0)=K^T(\omega,0)$ that guarantees a positive spectral function, defined by $-2\,  \mathrm{Im}(\sqrt{k^2-\omega^2} K^T)$. The same result could also be derived by direct resolution of \eqref{eq:symA}  for $k=0$.}
\begin{equation}
\sigma_x(\omega,0)=\sigma_y(\omega,0)=4 \kappa_N\,.
\label{eq:condzm}
\end{equation}
Therefore, the conductivity at zero momentum in any CFT holographic to a duality-invariant theory is a universal constant, independent of the frequency. Evidently, this is a consequence of the universal relation \eqref{eq:univresse}, as remarked in \cite{Herzog:2007ij,Myers:2010pk}. If duality symmetry is absent, Eq. \eqref{eq:condzm} may not necessarily hold --- see \cite{Myers:2010pk}.



For non-zero momentum $k$, the longitudinal and transverse conductivities $\sigma_x$ and $\sigma_y$ are no longer the same and possess an explicit frequency dependence, as already observed in Einstein-Maxwell theory \cite{Herzog:2007ij}. Although an exact analytical expression for the conductivities at any frequency and momentum in an arbitrary duality-invariant theory seems currently out of reach (it remains elusive even in the Einstein-Maxwell case), it is in fact possible to obtain explicit results in certain limits. For large frequencies, straightforward application of the WKB approximation shows that
\begin{align}
\label{eq:lclfy}
\sigma_x(\omega,k)&=4 \kappa_N \frac{\omega}{\sqrt{\omega^2-k^2}}\, , \quad \omega^2>>k^2 \,, \\
\sigma_y(\omega,k)&=4 \kappa_N \frac{\sqrt{\omega^2-k^2}}{\omega}\, , \quad \omega^2>>k^2 \,.
\label{eq:lclf}
\end{align} 
Therefore, the behaviour of conductivities for large frequencies is theory-independent. Besides, we note that they tend to the universal value \eqref{eq:condzm} as $\omega \rightarrow \infty$. On the other hand, in the limit of sufficiently small frequencies and momenta $\omega,k << r_0/\tilde{L}^2$, one may generalize the results in \cite{Policastro:2002se} for the retarded correlators to obtain
\begin{align}
\label{eq:lcsfy}
\sigma_x(\omega,k)&= \frac{4\kappa_N \omega^2}{\omega^2+ D^2 k^4}\,, \quad \omega,k << \frac{r_0}{\tilde{L}^2}  \\
\sigma_y(\omega,k)&= \frac{4 \kappa_N}{1+c\, k^2} \,, \quad \quad \, \, \omega,k << \frac{r_0}{\tilde{L}^2} \,,
\label{eq:lcsf}
\end{align}
where we have implicitly defined
\begin{align}
\label{eq:intdc}
D=\frac{\tilde{L}^2}{r_0}\int_0^{u_h} \frac{\mathrm{d}z}{\mathcal{B}(z)}\,,\quad  c=\frac{2\tilde{L}^4}{r_0^2}\int_0^{u_h}\mathrm{d}z \, \mathcal{S}(z) \int_z^{u_h} \frac{\mathrm{d}w}{\mathcal{B}(w)}   \,.
\end{align}
These expressions show that the conductivities for non-zero momenta will generically depend on both the gravitational background and the particular choice of duality-invariant theory (since this is the case for small frequencies and momenta). Also, a closer look at Eqs. \eqref{eq:lclfy} and \eqref{eq:lcsfy} reveals that the longitudinal conductivity undergoes a hydrodynamic-to-collisionless crossover\footnote{This hydrodynamic-to-collisionless crossover manifests in the longitudinal conductivity and not in the transverse one since the longitudinal correlator $C_{xx}$ is related via Eq. \eqref{eq:clong} to the density-density correlator $C_{tt}$, which is a clear probe of hydrodynamic behaviour \cite{Herzog:2007ij}.} as we go from small to large frequencies. This is signaled by the fact that Eq.  \eqref{eq:lcsfy} possesses a pole at $\omega=-i D k^2$ (which is precisely the dispersion relation of diffusion modes in the heat equation), while \eqref{eq:lclfy} presents a pole at $\omega=k$ (which is the dispersion relation for free particles). Moreover, as a consistency check of our results, we have verified that the expression for the diffusion constant that can be derived from the membrane paradigm  \cite{Kovtun:2003wp,Brigante:2007nu,Iqbal:2008by} coincides with our formula \eqref{eq:intdc} for $D$.



Away from the small/large frequency regimes, it appears to be challenging to obtain specific formulae for the conductivities. Despite that, a universal statement regarding their form in generic duality-invariant theories can be made by noticing that every traceless and symmetric tensor $\mathcal{T}_{\mu \nu}$ built from algebraic combinations (\emph{i.e.} with no covariant derivatives) of the curvature of \eqref{eq:bhbackground} vanishes  identically when evaluated on the GR AdS black brane solution:
\begin{equation}
\mathcal{T}_{\mu \nu}\vert_{N=1, f=1-u^3}=0\,.
\label{eq:tmunueins}
\end{equation}
The explicit proof of this result is given in the appendix. Therefore, the retarded correlators $C_{ab}$ and the conductivities $\sigma_x(\omega,k)$ and $\sigma_y(\omega,k)$  will coincide with those of Einstein-Maxwell theory for any duality-invariant theory with no covariant derivatives of the curvature in $\mathcal{T}_{\mu \nu}$ as long as the GR AdS black brane background is considered. It is important to note that taking the background to be that of GR does not imply that $\mathcal{L}_{\mathrm{grav}}^{\mathrm{high}}=0$ in \eqref{eq:dualgen}. Indeed, there exist myriads of higher-order gravities which do not correct the GR AdS black brane solution (\emph{e.g.} the well-known $f(R)$ gravities \cite{Buchdahl:1970ynr,Sotiriou:2008rp}). Therefore, one may interpret duality invariance as a very powerful tool to constrain observables to have a simple and fixed expression: that of Einstein-Maxwell theory.

If higher-curvature terms correct the GR solution, the retarded correlator, and hence the associated conductivities, will generically differ from\footnote{If one insists on working with the GR solution, the lowest-order choice for $\mathcal{T}_{\mu \nu}$ which is nonzero on the AdS black brane solution of GR is the traceless part of $\nabla_\mu W_{\alpha \beta \rho \sigma} \nabla_\nu W^{\alpha \beta \rho\sigma}$, where $W_{\alpha \beta \rho \sigma}$ stands for the Weyl tensor. This term would in principle modify the conductivities of the dual CFT. Nevertheless, such a term is of order $\mathcal{O}(L^8)$ and there are other higher-curvature terms of less order that already modify the gravitational background.} those of Einstein-Maxwell theory. In such a case, the subsequent charge transport will no longer be independent of the choice of duality-invariant theory. However, in spite of this lack of universality, if $\mathcal{T}_{\mu \nu}$ contains no covariant derivatives of the curvature it turns out that duality invariance forces corrections with respect to the case $\mathcal{T}_{\mu \nu}=0$ to be highly suppressed, since Eq. \eqref{eq:tmunueins} implies that $\mathcal{T}_{\mu \nu}\vert_{N,f}$ is subleading with respect to the leading-order corrections in the gravitational background. Therefore, corrections associated to the specific choice of $\mathcal{T}_{\mu \nu}$ (without covariant derivatives of the curvature) are subleading with respect to those arising from the choice of the gravitational background (or equivalently, of $\mathcal{L}^{\mathrm{high}}_{\mathrm{grav}}$).
\begin{figure}[h]
\centering
\includegraphics[scale=0.36,trim={0.35cm 0 0 0},clip]{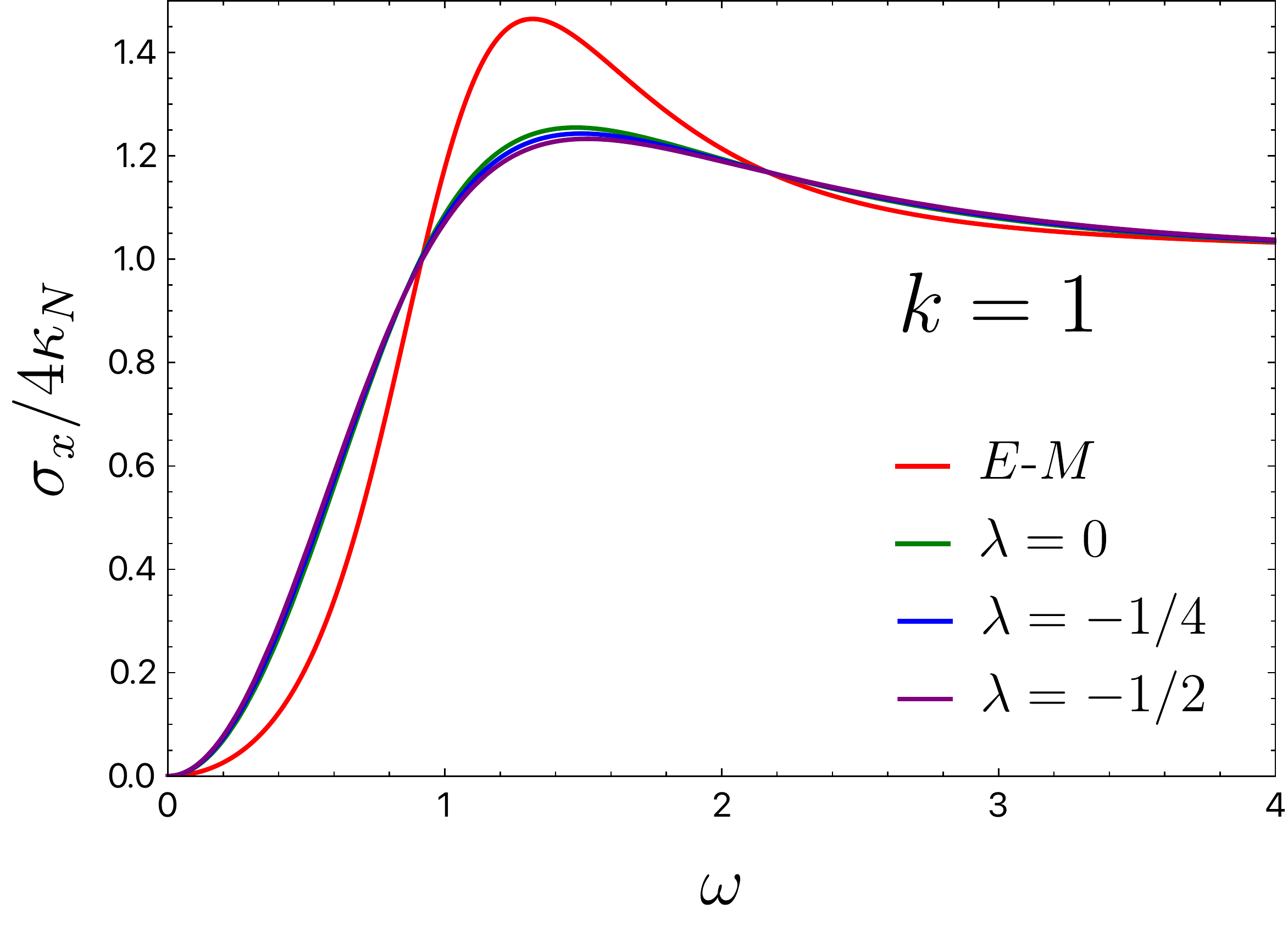}
\includegraphics[scale=0.36,trim={0.35cm 0 0 0},clip]{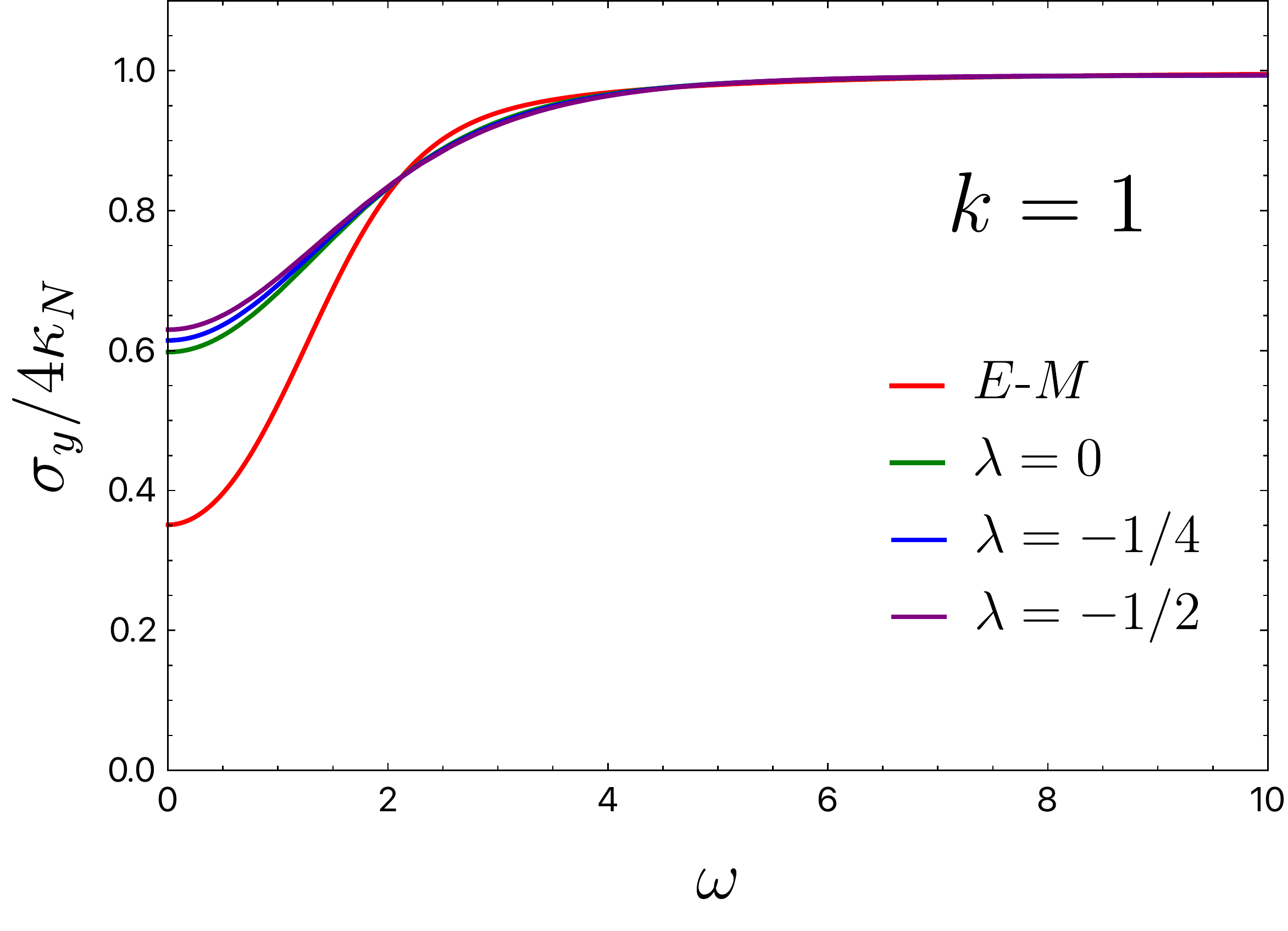}
\caption{Longitudinal (above) and transverse (below) conductivities in units of $L^2/r_0=1$ for Einstein-Maxwell ($E$-$M$) theory and for an ECG background. We have picked $\mu=1/10$, $k=1$  and several values of $\lambda$.}
\label{fig:cond}
\end{figure}



\textbf{Holographic conductivities in an explicit example.}
Now we illustrate the previous aspects with the simplest non-trivial choices for $\mathcal{T}_{\mu \nu}$ and $\mathcal{L}_{\mathrm{grav}}^{\mathrm{high}}$. Regarding $\mathcal{T}_{\mu \nu}$, this corresponds to
\begin{equation}
\mathcal{T}_{\mu \nu}=\lambda L^2 \hat{R}_{\mu \nu}\,,
\label{eq:dualchoice}
\end{equation}
where $\hat{R}_{\mu \nu}$ denotes the traceless part of the Ricci tensor and $\lambda$ is a dimensionless coupling. Demanding $\mathcal{T}_{\mu \nu}$ to respect the Weak Gravity Conjecture \cite{Arkani-Hamed:2006emk} and causality \cite{Myers:2010pk}, we find that the acceptable range for $\lambda$ is $0 \geq \lambda \gtrsim -0.50105$ (see appendix) for the specific choice of $\mathcal{L}_{\mathrm{grav}}^{\mathrm{high}}$ we are about to make.

To pick a suitable $\mathcal{L}_{\mathrm{grav}}^{\mathrm{high}}$ one needs to consider gravitational theories which contain at least terms of cubic order in the curvature, since quadratic terms do not correct the (four-dimensional) GR AdS black brane solution. Among this class of theories, there is a unique subset admitting black brane solutions \eqref{eq:bhbackground} with $N(u)=1$ and second-order equation for $f$. All such theories are equivalent on ans\"atze of the form \eqref{eq:bhbackground}, so it is enough to select a convenient representative. We will choose it to be Einsteinian Cubic Gravity (ECG) \cite{Bueno:2016xff,Bueno:2019ltp}, whose higher order terms have the following form:
\begin{align}
\nonumber
\hspace{-0.1cm}-8 \mathcal{L}_{\mathrm{grav}}^{\mathrm{high}}&=\mu L^4 \left[ 12 R_{\mu}{}^\rho{}_\nu{}^\sigma R_\rho{}^\gamma{}_\sigma{}^\delta R_\gamma{}^\mu{}_\delta{}^\nu+8 R_{\mu \nu} R^{\nu \rho} R_{\rho}{}^\mu\right.\\ &\left. +R_{\mu \nu}{}^{\rho \sigma} R_{\rho \sigma}{}^{\gamma \delta} R_{\gamma \delta}{}^{\mu \nu}-12 R_{\mu \nu \rho \sigma} R^{\mu \rho} R^{\nu \sigma}  \right]\,,
\label{eq:ecg}
\end{align} 
where $\mu$ is a dimensionless coupling. The equation of motion for $f(u)$, though second-order, is too complicated to be solved analytically for generic $\mu$, so we will resort to numeric methods (details are given in the appendix). We will pick $\mu$ to be within the range $0 < \mu < 4/27$, since this ensures the existence of both a unique stable vacuum and black hole solutions \cite{Bueno:2018xqc}. 

In Fig. \ref{fig:cond} we present the longitudinal and transverse conductivities we get for Einstein-Maxwell theory and for the choices \eqref{eq:dualchoice} and \eqref{eq:ecg}. We have  set $\mu=1/10$, $L^2 k/r_0=1$ and $\lambda=0,-1/2,-1/4$, since the qualitative behaviour of the conductivities turns out to replicate for any $0 < \mu <4/27$ (approaching of course the Einstein-Maxwell case as $\mu \to 0$) and $k$ (approaching the constant universal value \eqref{eq:condzm} as $k \rightarrow 0$). By direct inspection of the graphs we check that corrections associated to the specific choice of $\lambda$ --- \emph{i.e.} of $\mathcal{T}_{\mu \nu}$ --- are clearly subleading with respect to those arising from the choice of the gravitational background (characterized, in this case, by the parameter $\mu$).



\textbf{Final comments.} We have examined various universal aspects of the holographic quantum critical transport associated to duality-invariant theories. In the first place, we have explicitly checked that the conductivity at zero momentum is a universal constant for all these theories. Next we have obtained the expressions for the conductivities in the limit of large frequencies and for small frequencies and momenta in every CFT holographic to a duality-invariant theory. From their form in this latter regime, we have concluded that conductivities at non-zero momentum generically depend on both the gravitational background and the theory under study.


Despite that, we have proven that the conductivities in CFTs associated to duality-invariant theories which do not couple covariant derivatives of the curvature to gauge field strengths display two universal features. First, we have shown that, as long as a GR background is chosen, conductivities are universal and equal to those of Einstein-Maxwell theory for any frequency and momentum. Secondly, when the gravitational background is corrected by higher-curvature terms, we have proven that conductivities get modified in such a way that contributions from nonminimal couplings of the gauge field to gravity are subleading.





In another vein, there are several directions that would be interesting to address. Firstly, one could study other correlators of CFTs holographic to duality-invariant theories. For instance,  consider the Euclidean correlators $\langle J_a J_b \rangle_E$ and $\langle T_{ab} J_c J_d \rangle_E$ at zero temperature, where $T_{ab}$ is the stress-energy tensor. Conformal symmetry fixes the form of such correlators as follows \cite{Osborn:1993cr,Erdmenger:1996yc}: 
\begingroup
\begin{align}
   \langle J_a (x_1) J_b(x_2) \rangle_E&=\frac{C_J}{\vert x_{12} \vert^{4}} \mathcal{I}_{ab}\,, \\
   \langle T_{ab} (x_1) J_c (x_2) J_d(x_3) \rangle_E &=\frac{f_{abcd}(C_J,a_2)}{\vert x_{12} \vert^3 \vert x_{13} \vert^3 \vert x_{23} \vert}\,,
\end{align}
\endgroup
where $\mathcal{I}_{ab}$ and $f_{abcd}(C_J,a_2)$ are fixed tensorial structures, $x_{mn} =x_m-x_n$, $C_J$ is the current central charge and $a_2$ is a parameter that controls, together with $C_J$, the energy flux measured at infinity after the insertion of a current operator \cite{Hofman:2008ar}. The holographic expressions for $C_J$ and $a_2$ for the most general effective four-derivative theory were presented in \cite{Hofman:2008ar,Cano:2022ord}. Applying their results to the choice \eqref{eq:dualchoice} and $\mathcal{L}_{\mathrm{grav}}^{\mathrm{high}}=0$, one finds $a_2=0$ and that $C_J$ takes its Einstein-Maxwell value, so the Euclidean correlators at zero temperature are not modified to the fourth-derivative order.   This is another manifestation of the strength of duality invariance to constrain the form of correlators to be those of Einstein-Maxwell.

Secondly, it would be intriguing to extend our results to systems with chemical potentials (\emph{i.e.} with non-vanishing expectation values of global conserved charges in the equilibrium state). This is carried out by considering linear fluctuations of the vector field on top of a fixed charged gravitational background with a non-zero background electromagnetic field, as in \cite{Hartnoll:2007ai,Hartnoll:2007ih,Goldstein:2009cv,Blake:2014yla,Guo:2017bru,Wang:2018hwg,Kiczek:2020ngg}. 

Finally, one could also examine higher-point correlators. Indeed, it is natural to wonder what constraints duality invariance could impose on generic (current) $n$-point correlators. This would require the construction of (all) duality-invariant nonminimal extensions of Einstein-Maxwell theory of arbitrary order in $F_{\mu \nu}$, which remains as an outstanding open problem.

\vspace{1mm}
\noindent
\textbf{Acknowledgements}
\vspace{1mm}

\noindent
We would like to thank Pablo A. Cano for very useful and enlightening discussions.
 Á. J. M. was supported by a postdoctoral fellowship from the INFN, Bando 23590. D.S. acknowledges partial support  of
the Spanish MICINN/FEDER grant PID2021-125700NB-C21 and the Basque Government Grant IT-979-16.




\onecolumngrid  \vspace{0.8cm} 
\begin{center}  
{\Large\bf Appendices} 
\end{center} 
\appendix 

 \vspace{-0.25cm} 

\section{Traceless and symmetric two-tensors evaluated on black brane backgrounds}

Assume the following black brane ansatz for the metric:
\begin{equation}
    \diff s^2=\frac{r_0^2}{\tilde{L}^2 u^2}\left (-N^2(u)f(u) \diff t^2+\diff x^2+\diff y^2 \right )+\frac{\tilde{L}^2}{u^2 f(u)} \diff u^2\,,
    \label{eq:appbhbackground}
\end{equation}
for certain functions $N$ and $f$. We set the asymptotic (boundary) region to be at $u \rightarrow 0$, so that $\lim_{u \rightarrow 0}f=\lim_{u \rightarrow 0}N=1$, while the horizon (assuming it exists) is at $u=u_h$. Here we prove the following two features of traceless symmetric tensors $\mathcal{T}_{\mu \nu}$ evaluated on a black brane ansatz \eqref{eq:appbhbackground}:
\begin{itemize}
    \item Any $\mathcal{T}_{\mu \nu}$ built out from contractions of the curvature tensors and covariant derivatives thereof satisfies
    \begin{equation}
    \label{eq:approof1}
     \left. \mathcal{T}_{\mu}{}^{\nu}  \right \vert_{N,f}=2(\theta+\varphi)\delta_\mu{}^t \delta_t{}^\nu+2(\theta-\varphi)\delta_\mu{}^u \delta_u{}^\nu-\theta\tensor{\delta}{_\mu^\nu}\,,
    \end{equation}
    for certain functions $\theta$ and $\varphi$ of the coordinate $u$.
    \item Any $\mathcal{T}_{\mu \nu}$ built out from algebraic contractions of the curvature tensors (without covariant derivatives) vanishes when evaluated on the GR black brane solution given by $N=1$ and $f=1-u^3$:
    \begin{equation}
    \left. \mathcal{T}_{\mu \nu} \right \vert_{N=1,f=1-u^3}=0\,.
    \label{eq:approof2}
\end{equation} 
\end{itemize}
Let us start by showing the first property. Using the Ricci decomposition of the Riemann tensor, $\mathcal{T}_{\mu \nu}$ may be entirely expressed in terms of the Ricci scalar $R$, the traceless part $\hat{R}_{\mu \nu}$ of the Ricci tensor and the Weyl tensor $W_{\mu \nu}{}^{\rho \sigma}$. If we define the mutually orthogonal projectors $\tau_{\mu}{}^{\nu}=\tensor{\delta}{_\mu^t}\tensor{\delta}{_t^\nu}$, $\rho_{\mu}{}^{\nu}=\tensor{\delta}{_\mu^u}\tensor{\delta}{_u^\nu}$ and $\sigma_{\mu}{}^{\nu}=\tensor{\delta}{_\mu^x}\tensor{\delta}{_x^\nu}+\tensor{\delta}{_\mu^y}\tensor{\delta}{_y^\nu}$ onto the $t$, $u$ and $(x,y)$ directions respectively, then the traceless Ricci tensor and the Weyl tensor on gravitational configurations \eqref{eq:appbhbackground} read
\begin{align}
\label{eq:wricapp}
\left.  W_{\mu \nu}{}^{\rho \sigma}\right \vert_{N,f}&=w \left [ \zeta_{[\mu}{}^{[\rho} \zeta_{\nu]}{}^{\sigma]} -\zeta_{[\mu}{}^{[\rho} \sigma_{\nu]}{}^{\sigma]} +\sigma_{[\mu}{}^{[\rho} \sigma_{\nu]}{}^{\sigma]}    \right]\,, \quad \left.\hat{R}_{\mu \nu}\right \vert_{N,f}=(X+Y) \tau_{\mu \nu}+
  (X-Y) \rho_{\mu \nu}-X \sigma_{\mu \nu}\,,\\
  \label{eq:wxy}
  w=-&\frac{u^2(3f'N'+Nf''+2 f N'')}{3 \tilde{L}^2 N}\,, \quad X=-\frac{u(3uf'N'+N(u f''-2 f')-2f(N'-u N'')}{4 \tilde{L}^2 N}\, , \quad Y=\frac{u f N'}{\tilde{L}^2 N}\,,
\end{align}
where $\zeta_{\mu \nu}=\tau_{\mu \nu}+\rho_{\mu \nu}$. If $\mathcal{T}_{\mu \nu}$ is constructed from algebraic combinations of curvature tensors (with no covariant derivatives), then \eqref{eq:approof1} follows by direct inspection of \eqref{eq:wricapp}, which ensures that any contraction of curvature tensors will be expressed in terms of the projectors $\tau_{\mu \nu}$, $\rho_{\mu \nu}$ and $\sigma_{\mu \nu}$. Now,  assume that $\mathcal{T}_{\mu \nu}$ contains covariant derivatives of the curvature. We observe that
\begin{align}
    \nabla_{\mu} \tau_{\eta \lambda}&=\left (\frac{2}{u}-\frac{f'}{f}-\frac{2N'}{N} \right) \tau_{\mu (\eta} \, \delta_{\lambda)}^u\, , \quad \nabla_\mu \rho_{\eta \lambda}=-\nabla_{\mu} \tau_{\eta \lambda}-\frac{2}{u} \sigma_{\mu (\eta} \, \delta_{\lambda)}^u\, , \quad \nabla_\mu \sigma_{\eta \lambda}=-\nabla_\mu \tau_{\eta \lambda}-\nabla_\mu \rho_{\eta \lambda}\,,
    \\ 
    \nabla_\mu \delta_\nu^{u}&=\frac{u}{2 \tilde{L}^2 N}(u Nf'-2f(N-uN')) \tau_{\mu \nu}+\frac{u(2f+u f')}{2 \tilde{L^2}}\rho_{\mu \nu}-\frac{u f}{\tilde{L}^2}\sigma_{\mu \nu}\,.
\end{align}
Consequently, it turns out that any contraction of curvature tensors and covariant derivatives thereof will be entirely expressed in terms of the projectors $\tau_{\mu \nu}$, $\rho_{\mu \nu}$, $\sigma_{\mu \nu}$ and in terms of $\delta_\rho^u$. Moreover, since each of the terms composing $\mathcal{T}_{\mu \nu}$ must possess an even number of covariant derivatives (otherwise, the number of indices does not match), we can always pair up tensors $\delta_\mu^u$ into projectors $\rho_{\mu \nu}$. Hence we deduce that
\begin{equation}
    \left. \mathcal{T}_{\mu \nu} \right \vert_{N,f}=F_1 \tau_{\mu \nu}+F_2 \rho_{\mu \nu}+F_3 \sigma_{\mu \nu}\,.
\end{equation}
Demanding $\left. \mathcal{T}_{\mu \nu} \right \vert_{N,f}$ to be traceless, we arrive at \eqref{eq:approof1}.

Regarding the proof of \eqref{eq:approof2}, observe that \eqref{eq:wricapp} and \eqref{eq:wxy} imply that $\left. \hat R_{\mu \nu}\right \vert_{N=1,f=1-u^3}=0$. Hence all potential non-trivial contributions to $\mathcal{T}_{\mu \nu}$ may only come from terms containing just Weyl tensors. However, from \eqref{eq:wricapp} we realize that the expression for the Weyl tensor is symmetric under the exchange of $\zeta_{\mu}{}^\nu$ and $\sigma_{\mu}{}^\nu$. Therefore, any contraction of Weyl tensors must be symmetric under this exchange, so that
\begin{equation}
    \mathcal{W}_{\mu \nu}^{(n)}=c_n ( \zeta_{\mu \nu}+\sigma_{\mu \nu})=c_n g_{\mu \nu}\,,
\end{equation}
where $\mathcal{W}_{\mu \nu}^{(n)}$ denotes any term formed by $n$ Weyl tensors. The traceless part of any $\mathcal{W}_{\mu \nu}^{(n)}$ is clearly zero and thus we arrive at \eqref{eq:approof2}. Note that \eqref{eq:approof2} is no longer true if $\mathcal{T}_{\mu \nu}$ possesses covariant derivatives of the curvature. An explicit counterexample is provided by the traceless part of $\nabla_\mu W_{\alpha \beta \eta \lambda} \nabla_\nu W^{\alpha \beta \eta \lambda}$, which is non-zero on the GR black brane solution $N=1$ and $f=1-u^3$.

\section{Holographic Prescription for Retarded Correlators}

Let us consider a generic duality-invariant theory with nonminimal couplings to gravity which is quadratic in the Maxwell field strength. According to \cite{Cano:2021hje}, it can be written as follows:
\begin{equation}
 I=\kappa_N \int \diff^4 x \sqrt{\vert g \vert} \left[R+\frac{6}{L^2}-\chi^{\mu \nu \rho \sigma} F_{\mu \nu} F_{\rho\sigma}+\mathcal{L}_{\mathrm{grav}}^{\mathrm{high}}\right]\,,
    \label{eq:appdualgen}
\end{equation}
where $\kappa_N=(16 \pi G)^{-1}$, $\mathcal{L}_{\mathrm{grav}}^{\mathrm{high}}$ stands for purely-gravitational higher-curvature terms and $\chi^{\mu \nu \rho \sigma}$ is given by
\begin{align}
\label{eq:appd1}
    \tensor{\chi}{_{\mu \nu}^{\rho \sigma}}&= \tensor{\delta}{_{[\mu}^{[\rho}} \tensor{\delta}{_{\nu]}^{\sigma]}}+\tensor{\Theta}{_{\mu \nu}^{\rho \sigma}}+\sum_{n=1}^\infty b_n \tensor{\Theta}{^{2n}_{\mu \nu}^{\rho \sigma}}\,,\\ \tensor{\Theta}{^{2n}_{\mu \nu}^{\rho \sigma}}&=\tensor{\Theta}{_{\mu \nu}^{\alpha_1 \alpha_2}}\tensor{\Theta}{_{\alpha_1 \alpha_2}^{\alpha_3 \alpha_4}} \cdots \tensor{\Theta}{_{\alpha_{4n-3} \alpha_{4n-2}}^{\rho \sigma}}\,, \quad  \tensor{\Theta}{_{\mu \nu}^{\rho \sigma}}=\tensor{\mathcal{T}}{_{[\mu}^{[\rho}} \tensor{\delta}{_{\nu]}^{\sigma]}}\,,
    \label{eq:appd2}
    \end{align}
where $\mathcal{T}_{\mu \nu}$ is a traceless and symmetric tensor constructed from the curvature tensor and covariant derivatives thereof. 

Consider now fluctuations of the bulk vector field $A_\mu$ on top of the neutral gravitational background \eqref{eq:appbhbackground}. After integration by parts, the piece $I_A$ of the action \eqref{eq:appdualgen} which is quadratic in the vector field reads
\begin{equation}
    I_A=-4\kappa_N \int \diff^4 x\,  \partial_\mu\left (  \sqrt{\vert g \vert} \chi^{\mu \nu \rho \sigma} A_{\nu} \partial_{\rho}A_\sigma \right )+2\kappa_N \int \diff^4 x \sqrt{\vert g \vert }\, \nabla_\mu\left ( \chi^{\mu \nu \rho \sigma} F_{\rho \sigma} \right) A_\nu \,.
\end{equation}
If we assume that  $A_\mu$ solves its field equations $\nabla_\mu\left ( \chi^{\mu \nu \rho \sigma} F_{\rho \sigma} \right)=0$, then
\begin{equation}
     I_A=-4 \kappa_N \int \diff^4 x\,  \partial_\mu\left (  \sqrt{\vert g \vert} \chi^{\mu \nu \rho \sigma} A_{\nu} \partial_{\rho}A_\sigma \right )\,.
     \label{eq:appstokes}
\end{equation}
Taking into account Eqs. \eqref{eq:appd2} and \eqref{eq:approof1}, one may derive that $\chi_{\mu \nu}{}^{\rho \sigma}$  evaluated on the metric \eqref{eq:appbhbackground} adopts the form
\begin{align}
\label{chiblack}
   \left.  \tensor{\chi}{_{\mu \nu}^{\rho \sigma}} \right \vert_{N,f}&=2 N \mathcal{B}\, \tensor{\tau}{_{[\mu}^{[\rho}} \tensor{\rho}{_{\nu]}^{\sigma]}}+2 f N \mathcal{S}\, \tensor{\tau}{_{[\mu}^{[\rho}} \tensor{\sigma}{_{\nu]}^{\sigma]}} +\frac{2}{f N \mathcal{S}}\, \tensor{\rho}{_{[\mu}^{[\rho}} \tensor{\sigma}{_{\nu]}^{\sigma]}} +\frac{1}{N \mathcal{B}}\,\tensor{\sigma}{_{[\mu}^{[\rho}} \tensor{\sigma}{_{\nu]}^{\sigma]}}\,, \\ N \mathcal{B}&=\theta+\sqrt{1+\theta^2}\, , \qquad  \qquad  \qquad f N \mathcal{S}=\varphi+\sqrt{1+\varphi^2}\,.
\end{align}
For instance, if one chooses $\mathcal{T}_{\mu \nu}=\lambda L^2 \hat{R}_{\mu \nu}$ (as in the main text), then $\theta=\lambda L^2 X$ and $\varphi=\lambda L^2 Y/2$, where $X$ and $Y$ were defined back at \eqref{eq:wxy}. Now, using the gauge $A_u=0$, direct application of Stokes' theorem on \eqref{eq:appstokes} reveals that
\begin{equation}\label{IA'}
    I_A=-4\kappa_N \int \diff^3 x   \left.  \sqrt{\vert g \vert} \left. \chi^{u a u b} \right \vert_{N,f} A_{a} A'_b \right \vert_{u=0}^{u=u_h} \,,
\end{equation}
where prime denotes differentiation with respect to the coordinate $u$ and where Latin indices refer to boundary coordinates $x^a=(t,x,y)$. If $p^a=(\omega, \mathbf{k})$, let us write $A_a$ in momentum space:
\begin{equation}\label{appFA}
 A_a (t,u,\mathbf{x})=\int  \frac{\diff^3 p}{(2 \pi)^3}  \, e^{-i \omega t +i \mathbf{k} \cdot \mathbf{x} }A_a (u,\omega,\mathbf{k})=\int  \frac{\diff^3 p}{(2 \pi)^3}  \, e^{-i \omega t +i \mathbf{k} \cdot \mathbf{x} } \tensor{M}{_{a}^{b}}(u,\omega,\mathbf{k}) A_b^0 (\omega,\mathbf{k})\,,
\end{equation}
where $A_a^0=A_a \vert_{u=0}$ and where we have implicitly defined the tensor $\tensor{M}{_{a}^{b}}$ satisfying that $\tensor{M}{_{a}^{b}}(u=0,\omega,\mathbf{k})=\tensor{\delta}{_a^b}$. Substituting \eqref{appFA} into \eqref{IA'} and performing the integration in the spacetime coordinates, we arrive to
\begin{equation}
    I_A=\int \frac{\diff^3 p}{(2 \pi)^3}A_a^0 (-p^c) \mathcal{F}^{a b}(u, p^c)   A_b^0 (p^c) \Big \vert_{u=0}^{u=u_h}\,,
    \label{eq:prevcorr}
\end{equation}
where we have defined
 \begin{equation}
     \mathcal{F}^{ab}=-\frac{2r_0 \kappa_N }{\tilde{L}^2} \hat{\chi}^{cd}\tensor{M}{_{c}^{a}}(u,-p^e)\tensor{{M'}}{_{d}^{b}}(u,p^e)\,, \quad
\hat{\chi}^{ab}=\left (\frac{1}{\mathcal{S}}-\mathcal{B} \right) \tensor{\delta}{_t^a} \tensor{\delta}{_t^b}+\frac{1}{\mathcal{S}}\eta^{ab} \,,
 \end{equation}
 with $\eta^{ab}$ the Minkowski metric and $\tensor{{M'}}{_{d}^{b}}=\partial_u \tensor{{M}}{_{d}^{b}}$. Following the holographic prescription put forward in \cite{Son:2002sd,Herzog:2007ij}, we find the following result for the current-current retarded correlator $C_{ab}$:
\begin{equation}
    C_{ab}=2 \left. \mathcal{F}_{ab}\right \vert_{u=0}=-\left. \frac{4 r_0 \kappa_N }{ \tilde{L}^2}M'_{ab}\right \vert_{u=0}\,,
    \label{eq:appcorr}
\end{equation}
where we used that $\mathcal{B}$ and $\mathcal{S}$ must satisfy that $\lim_{u \rightarrow 0} \mathcal{B}=\lim_{u \rightarrow 0} \mathcal{S}=1$. It is important to note that \eqref{eq:appcorr} holds for all holographic duality-invariant theories (even with higher order terms in $F_{\mu\nu}$). The reason for this lies in the fact that one just needs to work with terms up to quadratic order in $A_\mu$ to fully determine $C_{ab}$. On a different front, we note that a similar derivation of such retarded correlators for generic theories \eqref{eq:appdualgen} (not necessarily duality-invariant) but in the particular case of $N(u)=1$ can be found in \cite{Myers:2010pk}.


\section{Einsteinian Cubic Gravity}

Up to topological terms, every four-dimensional six-derivative higher-order gravity may be mapped, via perturbative field redefinitions of the metric, to the following theory:
\begin{equation}
\begin{split}
    I&=\kappa_N \int \mathrm{d}^4 x \sqrt{\vert g \vert}\left[R+\frac{6}{L^2}-\frac{\mu L^4}{8}\mathcal{L}^{\mathrm{ECG}} \right]\, ,
    \label{eq:appactionecg} \\
    \mathcal{L}^{\mathrm{ECG}}&= 12 R_{\mu}{}^\rho{}_\nu{}^\sigma R_\rho{}^\gamma{}_\sigma{}^\delta R_\gamma{}^\mu{}_\delta{}^\nu+R_{\mu \nu}{}^{\rho \sigma} R_{\rho \sigma}{}^{\gamma \delta} R_{\gamma \delta}{}^{\mu \nu}-12 R_{\mu \nu \rho \sigma} R^{\mu \rho} R^{\nu \sigma} +8 R_{\mu \nu} R^{\nu \rho} R_{\rho}{}^\mu\,.
\end{split}
\end{equation}
where $\mu$ a dimensionless coupling and where $\mathcal{L}^{\mathrm{ECG}}$ receives the name of Einsteinian Cubic Gravity \cite{Bueno:2016xff}. The theory \eqref{eq:appactionecg} possesses two remarkable properties: it only propagates the usual massless  and traceless graviton on maximally symmetric backgrounds \cite{Bueno:2016xff} and admits solutions of the form \eqref{eq:appbhbackground} with $N=1$ and such that the equation of motion of $f$ can be integrated into a second-order differential equation \cite{Hennigar:2017ego,Bueno:2017sui}.

 Before examining gravitational backgrounds of the form of \eqref{eq:appbhbackground}, first one needs to derive the vacua of the theory, \emph{i.e.} its maximally-symmetric solutions. In GR, this is somewhat trivial, since the cosmological constant scale $L$ coincides with the corresponding AdS radius $\tilde{L}$. However, this is not the case when higher-curvature terms are present \cite{Bueno:2016xff,Bueno:2017sui}. For the theory \eqref{eq:appactionecg}, the relation between $L$ and the background scale $\tilde{L}$ reads as follows:
\begin{equation}
    1-\gamma+\mu \gamma^3=0\, , \qquad \gamma=\frac{L^2}{\tilde{L}^2}\,.
\end{equation}
This equation admits real and positive solutions for $\gamma$ when $\mu \leq 4/27$. Demanding the existence of both a unique stable vacuum and black hole solutions further restricts $\mu$ to lie within the range $0 \leq \mu \leq 4/27$, so that $1 \leq \gamma \leq 3/2$ \cite{Bueno:2018xqc}. When $\mu=0$ the theory \eqref{eq:appactionecg} reduces to GR, while $\mu=4/27$ corresponds to the so-called critical Einsteinian Cubic Gravity \cite{Feng:2017tev}, for which the effective Newton constant diverges. Therefore, we will assume $0 < \mu < 4/27$.


Let us now study black brane solutions of \eqref{eq:appactionecg}. The equations of motion on such backgrounds  may be obtained through the so-called reduced action method \cite{Palais:1979rca,Deser:2003up,Bueno:2017sui}, according to which one just needs to evaluate the action \eqref{eq:appactionecg} on \eqref{eq:appbhbackground} and vary with respect to $f$ and $N$. For the case at hands, if $I_{N,f}$ denotes such reduced action, one finds that
\begin{equation}
   \left.  \frac{\delta I_{N,f} }{\delta f} \right \vert_{N=1}=0\,, \qquad \forall f\,,
   \label{eq:gqgcond}
\end{equation}
so there exist black brane solutions with $N=1$. The equation of motion for $f$ is obtained by varying $I_{N,f}$ with respect to $N$ and then setting $N=1$. Eq. \eqref{eq:gqgcond} ensures that the equation for $f$  can be 
 integrated into the following second-order equation:
\begin{equation}
 4-4(3-2 \gamma) u^3-4 \gamma f+4\gamma^3  \mu f^3+\gamma^3 \mu  u^3 (f')^3-3 \gamma^3 \mu u^3 f f' f''=0\,,
 \label{eq:numecg}
\end{equation}
where the arbitrary constant arising from the integration has been carefully chosen so that, asymptotically (\emph{i.e.} as $u \rightarrow 0$), $f=1-u^3+ \mathcal{O}(u^6)$. 

Eq. \eqref{eq:numecg} is a non-linear second-order differential equation for $f$ whose analytical resolution is extremely challenging. Nevertheless, one can always obtain an asymptotic expansion for $f$, which reads
\begin{equation}
    f=1-u^3+\frac{21 (\gamma-1)}{4 \gamma-6} u^6-\frac{(\gamma-1) (3763 \gamma-3786)}{4 (3-2 \gamma)^2} u^9+\frac{3 (\gamma-1)^2 (535373 \gamma-540192) }{8 (2 \gamma-3)^3}u^{12}+ \mathcal{O}(u^{15})\,.
    \label{eq:fasymp}
\end{equation}
However, such asymptotic expansion does not provide an accurate description of the near-horizon profile for $f$, so it is necessary to solve \eqref{eq:numecg} numerically to capture both asymptotic and near-horizon behaviours.  

The differential equation \eqref{eq:numecg} becomes extraordinarily stiff as $u \rightarrow 0$ and one should proceed with utmost caution to extract a correct numerical solution.  In particular, as described in \cite{Bueno:2018xqc}, one should first demand $f$ to be regular around the horizon $u=u_h$:
\begin{equation}
    f(u)=4 \pi \frac{\tilde{L}^2}{r_0} T(u_h-u)+\sum_{n=2}^\infty a_n (u_h-u)^n\,,
\end{equation}
where $T$ is to be identified with the black brane temperature. Substituting this expression into \eqref{eq:numecg}, one fixes all coefficients $a_{n>2}$ in terms of $a_2$, which remains free. This is the parameter to adjust in order to guarantee the proper asymptotics for $f$. Through a shooting algorithm, 
 it is possible to find a unique value for $a_2$ which allows one to extend the solution from the horizon into the region where the asymptotic expansion \eqref{eq:fasymp} applies, thus finding the numerical solution for $f$ with the wanted physical properties.


 \section{Weak Gravity Conjecture and causality constraints}

Let us consider an exactly duality-invariant theory of the form \eqref{eq:appdualgen} with
\begin{equation}
    \mathcal{T}_{\mu \nu}=\lambda L^2 \hat{R}_{\mu \nu}\, , \quad \mathcal{L}_{\mathrm{grav}}^{\mathrm{high}}=-\frac{\mu L^4}{8} \mathcal{L}^{\rm ECG}\,,
    \label{eq:appdualwgc}
\end{equation}
where $\lambda$ is a dimensionless coupling. The goal of this appendix is to derive bounds for the dimensionless coupling $\lambda$ by demanding the theory defined by \eqref{eq:appdualwgc} to respect two physically-meaningful conditions: the Weak Gravity Conjecture and causality.
\if{}
\begin{equation}
    I=\kappa_N \int \mathrm{d}^4 x \sqrt{\vert g \vert} \left[R+ \frac{6}{L^2}- F^2- \lambda L^2 R_{\mu \nu} T_M^{\mu \nu}-\frac{\lambda^2 L^4}{4}\left ( \hat{R}_{\mu}{}^\alpha \hat{R}_\nu{}^\beta+ \hat{R}_{\mu \rho} \hat{R}^{\rho \alpha} \delta_{\beta}{}^\nu \right ) F^{\mu \nu} F_{\alpha \beta} \right]-\frac{\mu L^4}{8}\mathcal{L}^{\mathrm{ECG}}\,,
    \label{eq:theorydualsmall}
\end{equation}
where $\lambda$ is a dimensionless coupling $T_M^{\mu \nu}=F^{\mu \alpha} F^{\nu}{}_\alpha-\frac{1}{4} g^{\mu \nu}F^2$ the Maxwell stress-energy tensor. This theory corresponds to the $\mathcal{O}(L^2)$ truncation of an exactly duality-invariant theory of the form \eqref{eq:appdualgen} with $\mathcal{T}_{\mu \nu}=\lambda L^2 \hat{R}_{\mu \nu}$ ($\hat{R}_{\mu \nu}$ being the traceless Ricci tensor) and $\mathcal{L}_{\mathrm{grav}}^{\mathrm{high}}=\mathcal{O}(L^4)$. The goal of this appendix is to derive bounds for the dimensionless coupling $\lambda$ by demanding the theory \eqref{eq:theorydualsmall} to respect two physically-meaningful conditions: the Weak Gravity Conjecture and causality.
\fi

On the one hand, the Weak Gravity conjecture (WGC) was originally stated for asymptotically flat black holes \cite{Arkani-Hamed:2006emk,Harlow:2022gzl}. In its mild form, the WGC claims that the decay of an extremal black hole into smaller black holes should be possible, at least from the point of view of charge and energy conservation \cite{Cheung:2018cwt,Hamada:2018dde}. However, in AdS space such requirement is somewhat trivial, since perturbative (arbitrarily small) higher-order corrections cannot violate it \cite{Cremonini:2019wdk}. Instead, one could make use of the proposal put forward in \cite{Cheung:2018cwt}, according to which the corrections to the black hole entropy for any mass and charge ought to be positive for thermodynamically-stable black holes. This is a natural candidate for the generalization of the WGC to AdS spaces, since it has been shown that corrections to the near extremal entropy are related to corrections to the extremal mass \cite{Goon:2019faz,McPeak:2021tvu}. Therefore, this is the condition we will explore here. 


Focusing on black brane configurations of the type \eqref{eq:appbhbackground}, the argument requires to consider a complete charged black brane solution that takes into account the backreaction from the gauge field strength (in other words, we must find the appropriate corrections to the Reissner-Nordstr\"om solution). Imposing $F$ to have the same symmetries as \eqref{eq:appbhbackground}, we may write
\begin{equation}
 F=-\frac{r_0}{L}\Phi'\mathrm{d}t \wedge \mathrm{d}u+\frac{r_0^2}{L^3}P \, \mathrm{d}x \wedge \mathrm{d}y\,,
\end{equation}
where $\Phi=\Phi(u)$ is the electric potential and $P$ is a constant related to the magnetic charge. To derive strong constraints on $\lambda$ from the WGC conjecture, it suffices to consider the following  $\mathcal{O}(L^2)$ truncation  of \eqref{eq:appdualwgc}:
\begin{equation}
    I=\kappa_N \int \mathrm{d}^4 x \sqrt{\vert g \vert} \left[R+ \frac{6}{L^2}- F^2- \lambda L^2 R_{\mu \nu} T_M^{\mu \nu} \right]\,,
    \label{eq:theorydualsmall}
\end{equation}
where $T_M^{\mu \nu}=F_{\mu \alpha}F_{\nu}{}^\alpha-g_{\mu \nu} F^2/4$ is the Maxwell stress-energy tensor. Expand the solution in terms of $\lambda$:
\begin{equation}
    f=f_0+f_1 \lambda + \mathcal{O}(\lambda^2)\, , \quad  N=N_0+N_1 \lambda + \mathcal{O}(\lambda^2)\, , \quad  \Phi=\Phi_0+\Phi_1 \lambda + \mathcal{O}(\lambda^2)\, .
    \label{eq:appexpan}
\end{equation}
where the zeroth-order solution corresponds to the Reissner-Nordstr\"om one,
\begin{equation}
    f_0=1-u^3+ (Q^2+P^2)u^4\, ,\quad N_0=1 \, ,\quad \Phi_0=Q u\,,
\end{equation}
with $Q$ being a constant related to the electric charge. The computation of the leading-order corrections gives
\begin{equation}
    f_1=\frac{(Q^2+P^2) u^4}{20} \left (30 +  (14 (Q^2+P^2) u-15)u^3 \right )\, , \quad N_1=-\frac{u^4 (Q^2+P^2)}{4}\, , \quad \Phi_1=\frac{3 Q (Q^2+P^2) u^5}{20}\,.
\end{equation}
Observe that both $f_1$ and $N_1$ are invariant under rotations of $Q$ and $P$, as required by duality invariance. 

Let $u_h$ be the location of the outermost horizon. We express it as
\begin{equation}
    u_h=u_0+u_1 \lambda+\mathcal{O}(\lambda^2)\,.
\end{equation}
From the condition $f(u_h)=0$, one derives that 
\begin{equation}
    1-u_0^3+(Q^2+P^2) u_0^4=0\, , \quad u_1=\frac{u_0(16-u_0^3)(u_0^3-1)}{20(4-u_0^3)}\,.
\end{equation}
There exist positive solutions for $u_0$ only if and only if
\begin{equation}
    \frac{3}{4\sqrt[3]{4}} > Q^2+P^2>0\,.
    \label{eq:extcond}
\end{equation}
This imposes\footnote{The value $u_0=\sqrt[3]{4}$ corresponds to the extremal black brane and, in such a case, we observe that $u_1$ diverges. Therefore, the expansion in $\lambda$ is not reliable if we are arbitrarily close to extremality. Nevertheless, a near-extremal perturbative study of the black brane may be carried out if one remains in a regime for which $\vert \sqrt[3]{4}-u_0\vert >> \lambda$ (so that $\lambda u_1$ is still small compared to $u_0$).} $u_0 \in (1,\sqrt[3]{4})$, so that $u_1>0$. Since \eqref{eq:theorydualsmall} contains no higher-curvature terms, the entropy density $s$ associated to the black brane may be derived with the aid of the Bekenstein-Hawking formula, getting
\begin{equation}
    s=\frac{4 \pi \kappa_N r_0^2}{ L^2 u_h^2}=\frac{ 4 \pi \kappa_N r_0^2}{L^2 u_0^2}\left (1-\lambda \frac{(16-u_0^3)(u_0^3-1)}{10(4-u_0^3)} \right)+\mathcal{O}(\lambda^2)\,.
\end{equation}
For the correction to be positive, we must require\footnote{This constraint was obtained perturbatively and could not be reliable for large $\vert \lambda \vert$ (which would also imply the failure of the expansion \eqref{eq:appexpan}). Nevertheless, since causality will further restrict $\vert \lambda\vert $ to be less than $\sim 1/2$, we will not worry about further refinement.} $\lambda <0$. 

Let us now examine the constraints that may be derived  by requiring causality for the CFT dual to \eqref{eq:appdualwgc}. Following the analysis of \cite{Myers:2010pk}, one should guarantee that no superluminal modes of the vector field on top of a gravitational background of the form \eqref{eq:appbhbackground} are propagated, since this would signal causality violation in the holographic CFT. Such modes are absent when GR configurations are considered, but may appear after inclusion of higher-curvature terms. Therefore, we will explore which values of $\lambda$ ensure that causality is respected for the full theory \eqref{eq:appdualwgc}.


To this end, we have to study the equations of motion of $A_\mu$. In the gauge $A_u=0$, such equations were presented in the main text in momentum space. It was explained that the equations for $A_y$, $\mathcal{B} A_t'$ and $A_x'/\mathcal{S}$ are completely equivalent, so here it suffices  to analyze that of $A_y$.  Defining $r_0 z=\tilde{L}^2\int_0^u \mathcal{S}(u') \mathrm{d}u'$, this equation can be written as:
\begin{equation}
    -\frac{\mathrm{d}^2 A_y}{\mathrm{d} z^2}+ k^2 V(z) A_y= \omega^2 A_y\,, \quad V(z)=\frac{1}{\mathcal{B}(z) \mathcal{S}(z)}\,.
    \label{eq:ondasAy}
\end{equation}
By direct inspection of \eqref{eq:ondasAy}, we observe that the existence of a region for $z$ in which $V(z)>1$ will generically give rise to modes with superluminal propagation \cite{Brigante:2008gz,Buchel:2009tt}. Indeed, if $z_0$ is a point for which $V(z_0)>1$, then around this point the renormalizable modes (which need to satisfy $\omega^2-k^2 V(z_0)>0$) will be such that $\omega^2/k^2>1$. Therefore, we need to determine which values of $\lambda$ ensure that $V(z)\leq 1$ in the domain of $z$ corresponding to $u \in (0,u_h)$.



\begin{figure}[H]
\centering
\includegraphics[scale=0.4]{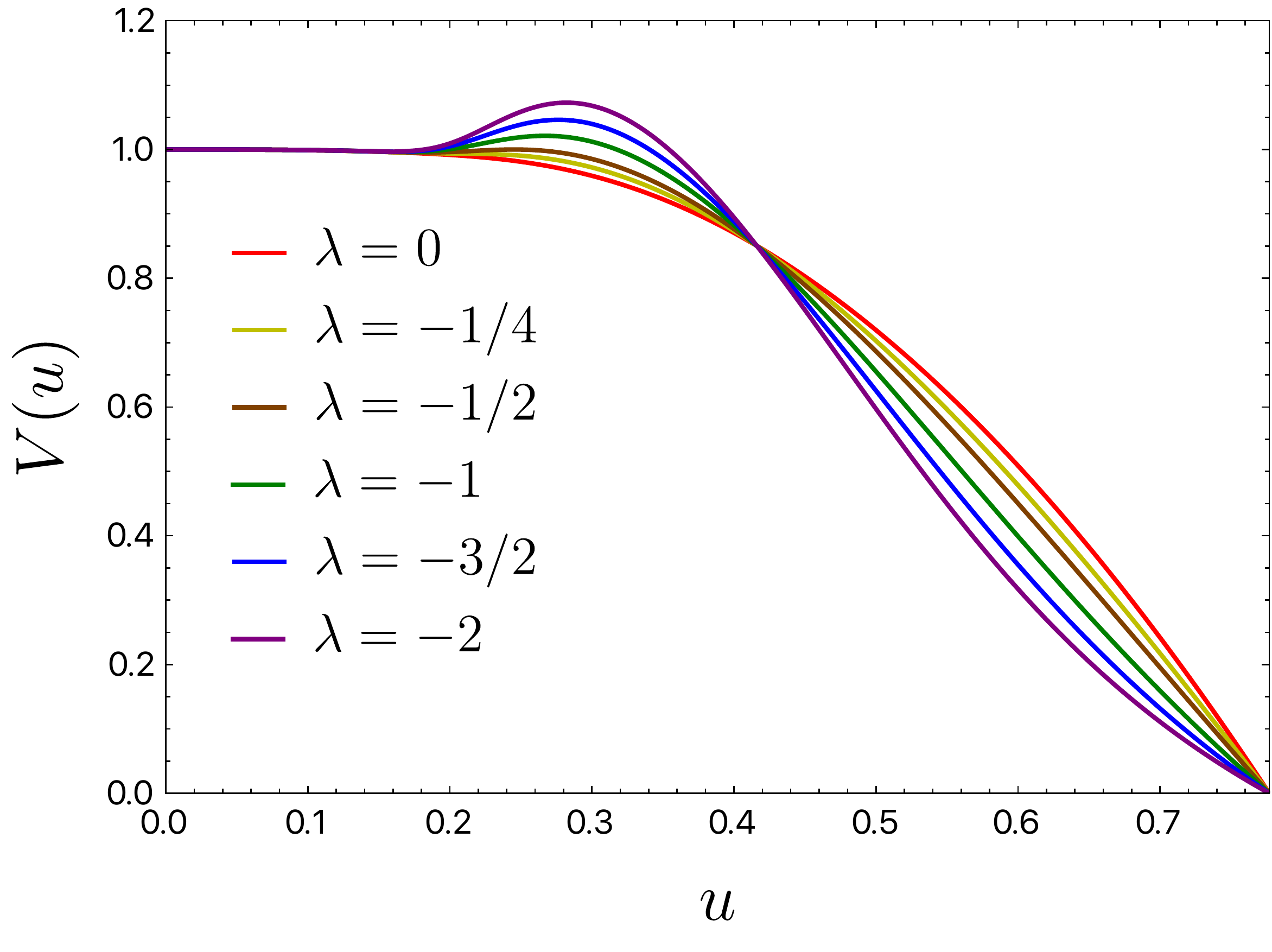}
\caption{Profile of $V(u)$. Here we have turned back to the use of the variable $u$ to make the representation clearer. We observe the appearance of a maximum that gets more and more pronounced as $\vert \lambda \vert$ increases.}
\label{fig:pots}
\end{figure}

The potential $V(z)$ depends on the parameter $\mu$ as well. We will fix it to be $\mu=1/10$, which is the value explored in the main text. Then, obtaining numerically the black brane background associated to \eqref{eq:appactionecg} (characterized by $N=1$ and the second order equation \eqref{eq:numecg} for $f$), in Fig. \ref{fig:pots} we present the profile for the potential for different values of $\lambda$. Only negative values of $\lambda$ are considered, since, as shown above, positive values are forbidden by the WGC conjecture. We observe that, as $\vert \lambda \vert$ increases, the function $V(z)$ develops a maximum which gets more and more peaked. In particular, we have checked that $V(z)\leq 1$ if
\begin{equation}
   0\geq \lambda \gtrsim -0.50105\,.
   \label{eq:lambdapermi}
\end{equation}
This is the range for $\lambda$ considered in the main text, obtained in the specific case $\mu=1/10$. Now, one might wonder whether additional conditions to further constrain the values of $\lambda$ could be found. For instance, one could explore if the energy flux measured at infinity after the local insertion of a current operator \cite{Hofman:2008ar} is positive in all directions \cite{Myers:2010pk} or study the existence of unstable modes of the vector field \cite{Myers:2007we}. However, none of these requirements provide additional constraints, since the energy flux is exactly the same as that of Einstein-Maxwell theory (as explained at the end of the body of the manuscript) and since the potential has no negative minima for the probed values of $\lambda$ \cite{Myers:2010pk}. 

\bibliographystyle{apsrev4-1}
\vspace{1cm}
\bibliography{Gravities.bib}

 \newcommand{\noop}[1]{}
\begin{thebibliography}{99}%
\makeatletter
\providecommand \@ifxundefined [1]{%
 \@ifx{#1\undefined}
}%
\providecommand \@ifnum [1]{%
 \ifnum #1\expandafter \@firstoftwo
 \else \expandafter \@secondoftwo
 \fi
}%
\providecommand \@ifx [1]{%
 \ifx #1\expandafter \@firstoftwo
 \else \expandafter \@secondoftwo
 \fi
}%
\providecommand \natexlab [1]{#1}%
\providecommand \enquote  [1]{``#1''}%
\providecommand \bibnamefont  [1]{#1}%
\providecommand \bibfnamefont [1]{#1}%
\providecommand \citenamefont [1]{#1}%
\providecommand \href@noop [0]{\@secondoftwo}%
\providecommand \href [0]{\begingroup \@sanitize@url \@href}%
\providecommand \@href[1]{\@@startlink{#1}\@@href}%
\providecommand \@@href[1]{\endgroup#1\@@endlink}%
\providecommand \@sanitize@url [0]{\catcode `\\12\catcode `\$12\catcode
  `\&12\catcode `\#12\catcode `\^12\catcode `\_12\catcode `\%12\relax}%
\providecommand \@@startlink[1]{}%
\providecommand \@@endlink[0]{}%
\providecommand \url  [0]{\begingroup\@sanitize@url \@url }%
\providecommand \@url [1]{\endgroup\@href {#1}{\urlprefix }}%
\providecommand \urlprefix  [0]{URL }%
\providecommand \Eprint [0]{\href }%
\providecommand \doibase [0]{http://dx.doi.org/}%
\providecommand \selectlanguage [0]{\@gobble}%
\providecommand \bibinfo  [0]{\@secondoftwo}%
\providecommand \bibfield  [0]{\@secondoftwo}%
\providecommand \translation [1]{[#1]}%
\providecommand \BibitemOpen [0]{}%
\providecommand \bibitemStop [0]{}%
\providecommand \bibitemNoStop [0]{.\EOS\space}%
\providecommand \EOS [0]{\spacefactor3000\relax}%
\providecommand \BibitemShut  [1]{\csname bibitem#1\endcsname}%
\let\auto@bib@innerbib\@empty
\bibitem [{\citenamefont {Maldacena}(1998)}]{Maldacena:1997re}%
  \BibitemOpen
  \bibfield  {author} {\bibinfo {author} {\bibfnamefont {J.~M.}\ \bibnamefont
  {Maldacena}},\ }\href {\doibase 10.1023/A:1026654312961} {\bibfield
  {journal} {\bibinfo  {journal} {Adv. Theor. Math. Phys.}\ }\textbf {\bibinfo
  {volume} {2}},\ \bibinfo {pages} {231} (\bibinfo {year} {1998})},\ \Eprint
  {http://arxiv.org/abs/hep-th/9711200} {arXiv:hep-th/9711200} \BibitemShut
  {NoStop}%
\bibitem [{\citenamefont {Gubser}\ \emph {et~al.}(1998)\citenamefont {Gubser},
  \citenamefont {Klebanov},\ and\ \citenamefont {Polyakov}}]{Gubser:1998bc}%
  \BibitemOpen
  \bibfield  {author} {\bibinfo {author} {\bibfnamefont {S.~S.}\ \bibnamefont
  {Gubser}}, \bibinfo {author} {\bibfnamefont {I.~R.}\ \bibnamefont
  {Klebanov}}, \ and\ \bibinfo {author} {\bibfnamefont {A.~M.}\ \bibnamefont
  {Polyakov}},\ }\href {\doibase 10.1016/S0370-2693(98)00377-3} {\bibfield
  {journal} {\bibinfo  {journal} {Phys. Lett. B}\ }\textbf {\bibinfo {volume}
  {428}},\ \bibinfo {pages} {105} (\bibinfo {year} {1998})},\ \Eprint
  {http://arxiv.org/abs/hep-th/9802109} {arXiv:hep-th/9802109} \BibitemShut
  {NoStop}%
\bibitem [{\citenamefont {Witten}(1998)}]{Witten:1998qj}%
  \BibitemOpen
  \bibfield  {author} {\bibinfo {author} {\bibfnamefont {E.}~\bibnamefont
  {Witten}},\ }\href {\doibase 10.4310/ATMP.1998.v2.n2.a2} {\bibfield
  {journal} {\bibinfo  {journal} {Adv. Theor. Math. Phys.}\ }\textbf {\bibinfo
  {volume} {2}},\ \bibinfo {pages} {253} (\bibinfo {year} {1998})},\ \Eprint
  {http://arxiv.org/abs/hep-th/9802150} {arXiv:hep-th/9802150} \BibitemShut
  {NoStop}%
\bibitem [{\citenamefont {Hartnoll}(2009)}]{Hartnoll:2009sz}%
  \BibitemOpen
  \bibfield  {author} {\bibinfo {author} {\bibfnamefont {S.~A.}\ \bibnamefont
  {Hartnoll}},\ }\href {\doibase 10.1088/0264-9381/26/22/224002} {\bibfield
  {journal} {\bibinfo  {journal} {Class. Quant. Grav.}\ }\textbf {\bibinfo
  {volume} {26}},\ \bibinfo {pages} {224002} (\bibinfo {year} {2009})},\
  \Eprint {http://arxiv.org/abs/0903.3246} {arXiv:0903.3246 [hep-th]}
  \BibitemShut {NoStop}%
\bibitem [{\citenamefont {Sachdev}(2011)}]{Sachdev:2010ch}%
  \BibitemOpen
  \bibfield  {author} {\bibinfo {author} {\bibfnamefont {S.}~\bibnamefont
  {Sachdev}},\ }\href {\doibase 10.1007/978-3-642-04864-7_9} {\bibfield
  {journal} {\bibinfo  {journal} {Lect. Notes Phys.}\ }\textbf {\bibinfo
  {volume} {828}},\ \bibinfo {pages} {273} (\bibinfo {year} {2011})},\ \Eprint
  {http://arxiv.org/abs/1002.2947} {arXiv:1002.2947 [hep-th]} \BibitemShut
  {NoStop}%
\bibitem [{\citenamefont {Zaanen}\ \emph {et~al.}(2015)\citenamefont {Zaanen},
  \citenamefont {Sun}, \citenamefont {Liu},\ and\ \citenamefont
  {Schalm}}]{Zaanen:2015oix}%
  \BibitemOpen
  \bibfield  {author} {\bibinfo {author} {\bibfnamefont {J.}~\bibnamefont
  {Zaanen}}, \bibinfo {author} {\bibfnamefont {Y.-W.}\ \bibnamefont {Sun}},
  \bibinfo {author} {\bibfnamefont {Y.}~\bibnamefont {Liu}}, \ and\ \bibinfo
  {author} {\bibfnamefont {K.}~\bibnamefont {Schalm}},\ }\href@noop {} {\emph
  {\bibinfo {title} {{Holographic Duality in Condensed Matter Physics}}}}\
  (\bibinfo  {publisher} {Cambridge Univ. Press},\ \bibinfo {year}
  {2015})\BibitemShut {NoStop}%
\bibitem [{\citenamefont {Hartnoll}\ \emph {et~al.}(2016)\citenamefont
  {Hartnoll}, \citenamefont {Lucas},\ and\ \citenamefont
  {Sachdev}}]{Hartnoll:2016apf}%
  \BibitemOpen
  \bibfield  {author} {\bibinfo {author} {\bibfnamefont {S.~A.}\ \bibnamefont
  {Hartnoll}}, \bibinfo {author} {\bibfnamefont {A.}~\bibnamefont {Lucas}}, \
  and\ \bibinfo {author} {\bibfnamefont {S.}~\bibnamefont {Sachdev}},\
  }\href@noop {} {\  (\bibinfo {year} {2016})},\ \Eprint
  {http://arxiv.org/abs/1612.07324} {arXiv:1612.07324 [hep-th]} \BibitemShut
  {NoStop}%
\bibitem [{\citenamefont {Baggioli}(2016)}]{Baggioli:2016rdj}%
  \BibitemOpen
  \bibfield  {author} {\bibinfo {author} {\bibfnamefont {M.}~\bibnamefont
  {Baggioli}},\ }\emph {\bibinfo {title} {{Gravity, holography and applications
  to condensed matter}}},\ \href@noop {} {Ph.D. thesis},\ \bibinfo  {school}
  {Barcelona, Autonoma U.} (\bibinfo {year} {2016}),\ \Eprint
  {http://arxiv.org/abs/1610.02681} {arXiv:1610.02681 [hep-th]} \BibitemShut
  {NoStop}%
\bibitem [{\citenamefont {Nastase}(2017)}]{nastase_2017}%
  \BibitemOpen
  \bibfield  {author} {\bibinfo {author} {\bibfnamefont {H.}~\bibnamefont
  {Nastase}},\ }\href {\doibase 10.1017/9781316847978} {\emph {\bibinfo {title}
  {String Theory Methods for Condensed Matter Physics}}}\ (\bibinfo
  {publisher} {Cambridge University Press},\ \bibinfo {year}
  {2017})\BibitemShut {NoStop}%
\bibitem [{\citenamefont {Hartnoll}\ \emph
  {et~al.}(2008{\natexlab{a}})\citenamefont {Hartnoll}, \citenamefont
  {Herzog},\ and\ \citenamefont {Horowitz}}]{Hartnoll:2008vx}%
  \BibitemOpen
  \bibfield  {author} {\bibinfo {author} {\bibfnamefont {S.~A.}\ \bibnamefont
  {Hartnoll}}, \bibinfo {author} {\bibfnamefont {C.~P.}\ \bibnamefont
  {Herzog}}, \ and\ \bibinfo {author} {\bibfnamefont {G.~T.}\ \bibnamefont
  {Horowitz}},\ }\href {\doibase 10.1103/PhysRevLett.101.031601} {\bibfield
  {journal} {\bibinfo  {journal} {Phys. Rev. Lett.}\ }\textbf {\bibinfo
  {volume} {101}},\ \bibinfo {pages} {031601} (\bibinfo {year}
  {2008}{\natexlab{a}})},\ \Eprint {http://arxiv.org/abs/0803.3295}
  {arXiv:0803.3295 [hep-th]} \BibitemShut {NoStop}%
\bibitem [{\citenamefont {Keski-Vakkuri}\ and\ \citenamefont
  {Kraus}(2008)}]{Keski-Vakkuri:2008ffv}%
  \BibitemOpen
  \bibfield  {author} {\bibinfo {author} {\bibfnamefont {E.}~\bibnamefont
  {Keski-Vakkuri}}\ and\ \bibinfo {author} {\bibfnamefont {P.}~\bibnamefont
  {Kraus}},\ }\href {\doibase 10.1088/1126-6708/2008/09/130} {\bibfield
  {journal} {\bibinfo  {journal} {JHEP}\ }\textbf {\bibinfo {volume} {09}},\
  \bibinfo {pages} {130} (\bibinfo {year} {2008})},\ \Eprint
  {http://arxiv.org/abs/0805.4643} {arXiv:0805.4643 [hep-th]} \BibitemShut
  {NoStop}%
\bibitem [{\citenamefont {Herzog}\ \emph {et~al.}(2009)\citenamefont {Herzog},
  \citenamefont {Kovtun},\ and\ \citenamefont {Son}}]{Herzog:2008he}%
  \BibitemOpen
  \bibfield  {author} {\bibinfo {author} {\bibfnamefont {C.~P.}\ \bibnamefont
  {Herzog}}, \bibinfo {author} {\bibfnamefont {P.~K.}\ \bibnamefont {Kovtun}},
  \ and\ \bibinfo {author} {\bibfnamefont {D.~T.}\ \bibnamefont {Son}},\ }\href
  {\doibase 10.1103/PhysRevD.79.066002} {\bibfield  {journal} {\bibinfo
  {journal} {Phys. Rev. D}\ }\textbf {\bibinfo {volume} {79}},\ \bibinfo
  {pages} {066002} (\bibinfo {year} {2009})},\ \Eprint
  {http://arxiv.org/abs/0809.4870} {arXiv:0809.4870 [hep-th]} \BibitemShut
  {NoStop}%
\bibitem [{\citenamefont {Hartnoll}\ \emph
  {et~al.}(2008{\natexlab{b}})\citenamefont {Hartnoll}, \citenamefont
  {Herzog},\ and\ \citenamefont {Horowitz}}]{Hartnoll:2008kx}%
  \BibitemOpen
  \bibfield  {author} {\bibinfo {author} {\bibfnamefont {S.~A.}\ \bibnamefont
  {Hartnoll}}, \bibinfo {author} {\bibfnamefont {C.~P.}\ \bibnamefont
  {Herzog}}, \ and\ \bibinfo {author} {\bibfnamefont {G.~T.}\ \bibnamefont
  {Horowitz}},\ }\href {\doibase 10.1088/1126-6708/2008/12/015} {\bibfield
  {journal} {\bibinfo  {journal} {JHEP}\ }\textbf {\bibinfo {volume} {12}},\
  \bibinfo {pages} {015} (\bibinfo {year} {2008}{\natexlab{b}})},\ \Eprint
  {http://arxiv.org/abs/0810.1563} {arXiv:0810.1563 [hep-th]} \BibitemShut
  {NoStop}%
\bibitem [{\citenamefont {Herzog}(2009)}]{Herzog:2009xv}%
  \BibitemOpen
  \bibfield  {author} {\bibinfo {author} {\bibfnamefont {C.~P.}\ \bibnamefont
  {Herzog}},\ }\href {\doibase 10.1088/1751-8113/42/34/343001} {\bibfield
  {journal} {\bibinfo  {journal} {J. Phys. A}\ }\textbf {\bibinfo {volume}
  {42}},\ \bibinfo {pages} {343001} (\bibinfo {year} {2009})},\ \Eprint
  {http://arxiv.org/abs/0904.1975} {arXiv:0904.1975 [hep-th]} \BibitemShut
  {NoStop}%
\bibitem [{\citenamefont {Bergman}\ \emph {et~al.}(2010)\citenamefont
  {Bergman}, \citenamefont {Jokela}, \citenamefont {Lifschytz},\ and\
  \citenamefont {Lippert}}]{Bergman:2010gm}%
  \BibitemOpen
  \bibfield  {author} {\bibinfo {author} {\bibfnamefont {O.}~\bibnamefont
  {Bergman}}, \bibinfo {author} {\bibfnamefont {N.}~\bibnamefont {Jokela}},
  \bibinfo {author} {\bibfnamefont {G.}~\bibnamefont {Lifschytz}}, \ and\
  \bibinfo {author} {\bibfnamefont {M.}~\bibnamefont {Lippert}},\ }\href
  {\doibase 10.1007/JHEP10(2010)063} {\bibfield  {journal} {\bibinfo  {journal}
  {JHEP}\ }\textbf {\bibinfo {volume} {10}},\ \bibinfo {pages} {063} (\bibinfo
  {year} {2010})},\ \Eprint {http://arxiv.org/abs/1003.4965} {arXiv:1003.4965
  [hep-th]} \BibitemShut {NoStop}%
\bibitem [{\citenamefont {Hartnoll}\ \emph {et~al.}(2021)\citenamefont
  {Hartnoll}, \citenamefont {Horowitz}, \citenamefont {Kruthoff},\ and\
  \citenamefont {Santos}}]{Hartnoll:2020fhc}%
  \BibitemOpen
  \bibfield  {author} {\bibinfo {author} {\bibfnamefont {S.~A.}\ \bibnamefont
  {Hartnoll}}, \bibinfo {author} {\bibfnamefont {G.~T.}\ \bibnamefont
  {Horowitz}}, \bibinfo {author} {\bibfnamefont {J.}~\bibnamefont {Kruthoff}},
  \ and\ \bibinfo {author} {\bibfnamefont {J.~E.}\ \bibnamefont {Santos}},\
  }\href {\doibase 10.21468/SciPostPhys.10.1.009} {\bibfield  {journal}
  {\bibinfo  {journal} {SciPost Phys.}\ }\textbf {\bibinfo {volume} {10}},\
  \bibinfo {pages} {009} (\bibinfo {year} {2021})},\ \Eprint
  {http://arxiv.org/abs/2008.12786} {arXiv:2008.12786 [hep-th]} \BibitemShut
  {NoStop}%
\bibitem [{\citenamefont {Arean}\ \emph {et~al.}(2021)\citenamefont {Arean},
  \citenamefont {Baggioli}, \citenamefont {Grieninger},\ and\ \citenamefont
  {Landsteiner}}]{Arean:2021tks}%
  \BibitemOpen
  \bibfield  {author} {\bibinfo {author} {\bibfnamefont {D.}~\bibnamefont
  {Arean}}, \bibinfo {author} {\bibfnamefont {M.}~\bibnamefont {Baggioli}},
  \bibinfo {author} {\bibfnamefont {S.}~\bibnamefont {Grieninger}}, \ and\
  \bibinfo {author} {\bibfnamefont {K.}~\bibnamefont {Landsteiner}},\ }\href
  {\doibase 10.1007/JHEP11(2021)206} {\bibfield  {journal} {\bibinfo  {journal}
  {JHEP}\ }\textbf {\bibinfo {volume} {11}},\ \bibinfo {pages} {206} (\bibinfo
  {year} {2021})},\ \Eprint {http://arxiv.org/abs/2107.08802} {arXiv:2107.08802
  [hep-th]} \BibitemShut {NoStop}%
\bibitem [{\citenamefont {Ammon}\ \emph {et~al.}(2022)\citenamefont {Ammon},
  \citenamefont {Arean}, \citenamefont {Baggioli}, \citenamefont {Gray},\ and\
  \citenamefont {Grieninger}}]{Ammon:2021pyz}%
  \BibitemOpen
  \bibfield  {author} {\bibinfo {author} {\bibfnamefont {M.}~\bibnamefont
  {Ammon}}, \bibinfo {author} {\bibfnamefont {D.}~\bibnamefont {Arean}},
  \bibinfo {author} {\bibfnamefont {M.}~\bibnamefont {Baggioli}}, \bibinfo
  {author} {\bibfnamefont {S.}~\bibnamefont {Gray}}, \ and\ \bibinfo {author}
  {\bibfnamefont {S.}~\bibnamefont {Grieninger}},\ }\href {\doibase
  10.1007/JHEP03(2022)015} {\bibfield  {journal} {\bibinfo  {journal} {JHEP}\
  }\textbf {\bibinfo {volume} {03}},\ \bibinfo {pages} {015} (\bibinfo {year}
  {2022})},\ \Eprint {http://arxiv.org/abs/2111.10305} {arXiv:2111.10305
  [hep-th]} \BibitemShut {NoStop}%
\bibitem [{\citenamefont {Sword}\ and\ \citenamefont
  {Vegh}(2022)}]{Sword:2021pfm}%
  \BibitemOpen
  \bibfield  {author} {\bibinfo {author} {\bibfnamefont {L.}~\bibnamefont
  {Sword}}\ and\ \bibinfo {author} {\bibfnamefont {D.}~\bibnamefont {Vegh}},\
  }\href {\doibase 10.1007/JHEP04(2022)135} {\bibfield  {journal} {\bibinfo
  {journal} {JHEP}\ }\textbf {\bibinfo {volume} {04}},\ \bibinfo {pages} {135}
  (\bibinfo {year} {2022})},\ \Eprint {http://arxiv.org/abs/2112.14177}
  {arXiv:2112.14177 [hep-th]} \BibitemShut {NoStop}%
\bibitem [{\citenamefont {Donos}\ and\ \citenamefont
  {Kailidis}(2022{\natexlab{a}})}]{Donos:2022www}%
  \BibitemOpen
  \bibfield  {author} {\bibinfo {author} {\bibfnamefont {A.}~\bibnamefont
  {Donos}}\ and\ \bibinfo {author} {\bibfnamefont {P.}~\bibnamefont
  {Kailidis}},\ }\href {\doibase 10.1007/JHEP11(2022)053} {\bibfield  {journal}
  {\bibinfo  {journal} {JHEP}\ }\textbf {\bibinfo {volume} {11}},\ \bibinfo
  {pages} {053} (\bibinfo {year} {2022}{\natexlab{a}})},\ \Eprint
  {http://arxiv.org/abs/2209.06893} {arXiv:2209.06893 [hep-th]} \BibitemShut
  {NoStop}%
\bibitem [{\citenamefont {Flory}\ \emph {et~al.}(2022)\citenamefont {Flory},
  \citenamefont {Grieninger},\ and\ \citenamefont
  {Morales-Tejera}}]{Flory:2022uzp}%
  \BibitemOpen
  \bibfield  {author} {\bibinfo {author} {\bibfnamefont {M.}~\bibnamefont
  {Flory}}, \bibinfo {author} {\bibfnamefont {S.}~\bibnamefont {Grieninger}}, \
  and\ \bibinfo {author} {\bibfnamefont {S.}~\bibnamefont {Morales-Tejera}},\
  }\href@noop {} {\  (\bibinfo {year} {2022})},\ \Eprint
  {http://arxiv.org/abs/2209.09251} {arXiv:2209.09251 [hep-th]} \BibitemShut
  {NoStop}%
\bibitem [{\citenamefont {Donos}\ and\ \citenamefont
  {Kailidis}(2022{\natexlab{b}})}]{Donos:2022qao}%
  \BibitemOpen
  \bibfield  {author} {\bibinfo {author} {\bibfnamefont {A.}~\bibnamefont
  {Donos}}\ and\ \bibinfo {author} {\bibfnamefont {P.}~\bibnamefont
  {Kailidis}},\ }\href {\doibase 10.1007/JHEP12(2022)028} {\bibfield  {journal}
  {\bibinfo  {journal} {JHEP}\ }\textbf {\bibinfo {volume} {12}},\ \bibinfo
  {pages} {028} (\bibinfo {year} {2022}{\natexlab{b}})},\ \Eprint
  {http://arxiv.org/abs/2210.06513} {arXiv:2210.06513 [hep-th]} \BibitemShut
  {NoStop}%
\bibitem [{\citenamefont {Buchel}\ \emph {et~al.}(2005)\citenamefont {Buchel},
  \citenamefont {Liu},\ and\ \citenamefont {Starinets}}]{Buchel:2004di}%
  \BibitemOpen
  \bibfield  {author} {\bibinfo {author} {\bibfnamefont {A.}~\bibnamefont
  {Buchel}}, \bibinfo {author} {\bibfnamefont {J.~T.}\ \bibnamefont {Liu}}, \
  and\ \bibinfo {author} {\bibfnamefont {A.~O.}\ \bibnamefont {Starinets}},\
  }\href {\doibase 10.1016/j.nuclphysb.2004.11.055} {\bibfield  {journal}
  {\bibinfo  {journal} {Nucl. Phys. B}\ }\textbf {\bibinfo {volume} {707}},\
  \bibinfo {pages} {56} (\bibinfo {year} {2005})},\ \Eprint
  {http://arxiv.org/abs/hep-th/0406264} {arXiv:hep-th/0406264} \BibitemShut
  {NoStop}%
\bibitem [{\citenamefont {Benincasa}\ and\ \citenamefont
  {Buchel}(2006)}]{Benincasa:2005qc}%
  \BibitemOpen
  \bibfield  {author} {\bibinfo {author} {\bibfnamefont {P.}~\bibnamefont
  {Benincasa}}\ and\ \bibinfo {author} {\bibfnamefont {A.}~\bibnamefont
  {Buchel}},\ }\href {\doibase 10.1088/1126-6708/2006/01/103} {\bibfield
  {journal} {\bibinfo  {journal} {JHEP}\ }\textbf {\bibinfo {volume} {01}},\
  \bibinfo {pages} {103} (\bibinfo {year} {2006})},\ \Eprint
  {http://arxiv.org/abs/hep-th/0510041} {arXiv:hep-th/0510041} \BibitemShut
  {NoStop}%
\bibitem [{\citenamefont {Myers}\ \emph {et~al.}(2009)\citenamefont {Myers},
  \citenamefont {Paulos},\ and\ \citenamefont {Sinha}}]{Myers:2008yi}%
  \BibitemOpen
  \bibfield  {author} {\bibinfo {author} {\bibfnamefont {R.~C.}\ \bibnamefont
  {Myers}}, \bibinfo {author} {\bibfnamefont {M.~F.}\ \bibnamefont {Paulos}}, \
  and\ \bibinfo {author} {\bibfnamefont {A.}~\bibnamefont {Sinha}},\ }\href
  {\doibase 10.1103/PhysRevD.79.041901} {\bibfield  {journal} {\bibinfo
  {journal} {Phys. Rev. D}\ }\textbf {\bibinfo {volume} {79}},\ \bibinfo
  {pages} {041901} (\bibinfo {year} {2009})},\ \Eprint
  {http://arxiv.org/abs/0806.2156} {arXiv:0806.2156 [hep-th]} \BibitemShut
  {NoStop}%
\bibitem [{\citenamefont {de~Boer}\ \emph {et~al.}(2010)\citenamefont
  {de~Boer}, \citenamefont {Kulaxizi},\ and\ \citenamefont
  {Parnachev}}]{deBoer:2009pn}%
  \BibitemOpen
  \bibfield  {author} {\bibinfo {author} {\bibfnamefont {J.}~\bibnamefont
  {de~Boer}}, \bibinfo {author} {\bibfnamefont {M.}~\bibnamefont {Kulaxizi}}, \
  and\ \bibinfo {author} {\bibfnamefont {A.}~\bibnamefont {Parnachev}},\ }\href
  {\doibase 10.1007/JHEP03(2010)087} {\bibfield  {journal} {\bibinfo  {journal}
  {JHEP}\ }\textbf {\bibinfo {volume} {03}},\ \bibinfo {pages} {087} (\bibinfo
  {year} {2010})},\ \Eprint {http://arxiv.org/abs/0910.5347} {arXiv:0910.5347
  [hep-th]} \BibitemShut {NoStop}%
\bibitem [{\citenamefont {Buchel}\ \emph {et~al.}(2010)\citenamefont {Buchel},
  \citenamefont {Escobedo}, \citenamefont {Myers}, \citenamefont {Paulos},
  \citenamefont {Sinha},\ and\ \citenamefont {Smolkin}}]{Buchel:2009sk}%
  \BibitemOpen
  \bibfield  {author} {\bibinfo {author} {\bibfnamefont {A.}~\bibnamefont
  {Buchel}}, \bibinfo {author} {\bibfnamefont {J.}~\bibnamefont {Escobedo}},
  \bibinfo {author} {\bibfnamefont {R.~C.}\ \bibnamefont {Myers}}, \bibinfo
  {author} {\bibfnamefont {M.~F.}\ \bibnamefont {Paulos}}, \bibinfo {author}
  {\bibfnamefont {A.}~\bibnamefont {Sinha}}, \ and\ \bibinfo {author}
  {\bibfnamefont {M.}~\bibnamefont {Smolkin}},\ }\href {\doibase
  10.1007/JHEP03(2010)111} {\bibfield  {journal} {\bibinfo  {journal} {JHEP}\
  }\textbf {\bibinfo {volume} {03}},\ \bibinfo {pages} {111} (\bibinfo {year}
  {2010})},\ \Eprint {http://arxiv.org/abs/0911.4257} {arXiv:0911.4257
  [hep-th]} \BibitemShut {NoStop}%
\bibitem [{\citenamefont {Myers}\ \emph {et~al.}(2010)\citenamefont {Myers},
  \citenamefont {Paulos},\ and\ \citenamefont {Sinha}}]{Myers:2010jv}%
  \BibitemOpen
  \bibfield  {author} {\bibinfo {author} {\bibfnamefont {R.~C.}\ \bibnamefont
  {Myers}}, \bibinfo {author} {\bibfnamefont {M.~F.}\ \bibnamefont {Paulos}}, \
  and\ \bibinfo {author} {\bibfnamefont {A.}~\bibnamefont {Sinha}},\ }\href
  {\doibase 10.1007/JHEP08(2010)035} {\bibfield  {journal} {\bibinfo  {journal}
  {JHEP}\ }\textbf {\bibinfo {volume} {08}},\ \bibinfo {pages} {035} (\bibinfo
  {year} {2010})},\ \Eprint {http://arxiv.org/abs/1004.2055} {arXiv:1004.2055
  [hep-th]} \BibitemShut {NoStop}%
\bibitem [{\citenamefont {Bueno}\ \emph {et~al.}(2018)\citenamefont {Bueno},
  \citenamefont {Cano},\ and\ \citenamefont {Ruip\'erez}}]{Bueno:2018xqc}%
  \BibitemOpen
  \bibfield  {author} {\bibinfo {author} {\bibfnamefont {P.}~\bibnamefont
  {Bueno}}, \bibinfo {author} {\bibfnamefont {P.~A.}\ \bibnamefont {Cano}}, \
  and\ \bibinfo {author} {\bibfnamefont {A.}~\bibnamefont {Ruip\'erez}},\
  }\href {\doibase 10.1007/JHEP03(2018)150} {\bibfield  {journal} {\bibinfo
  {journal} {JHEP}\ }\textbf {\bibinfo {volume} {03}},\ \bibinfo {pages} {150}
  (\bibinfo {year} {2018})},\ \Eprint {http://arxiv.org/abs/1802.00018}
  {arXiv:1802.00018 [hep-th]} \BibitemShut {NoStop}%
\bibitem [{\citenamefont {Cano}\ \emph {et~al.}(2022)\citenamefont {Cano},
  \citenamefont {Murcia}, \citenamefont {Rivadulla~S\'anchez},\ and\
  \citenamefont {Zhang}}]{Cano:2022ord}%
  \BibitemOpen
  \bibfield  {author} {\bibinfo {author} {\bibfnamefont {P.~A.}\ \bibnamefont
  {Cano}}, \bibinfo {author} {\bibfnamefont {{\'A}.~J.}\ \bibnamefont
  {Murcia}}, \bibinfo {author} {\bibfnamefont {A.}~\bibnamefont
  {Rivadulla~S\'anchez}}, \ and\ \bibinfo {author} {\bibfnamefont
  {X.}~\bibnamefont {Zhang}},\ }\href {\doibase 10.1007/JHEP07(2022)010}
  {\bibfield  {journal} {\bibinfo  {journal} {JHEP}\ }\textbf {\bibinfo
  {volume} {07}},\ \bibinfo {pages} {010} (\bibinfo {year} {2022})},\ \Eprint
  {http://arxiv.org/abs/2202.10473} {arXiv:2202.10473 [hep-th]} \BibitemShut
  {NoStop}%
\bibitem [{\citenamefont {Myers}\ and\ \citenamefont
  {Sinha}(2010)}]{Myers:2010xs}%
  \BibitemOpen
  \bibfield  {author} {\bibinfo {author} {\bibfnamefont {R.~C.}\ \bibnamefont
  {Myers}}\ and\ \bibinfo {author} {\bibfnamefont {A.}~\bibnamefont {Sinha}},\
  }\href {\doibase 10.1103/PhysRevD.82.046006} {\bibfield  {journal} {\bibinfo
  {journal} {Phys. Rev.}\ }\textbf {\bibinfo {volume} {D82}},\ \bibinfo {pages}
  {046006} (\bibinfo {year} {2010})},\ \Eprint {http://arxiv.org/abs/1006.1263}
  {arXiv:1006.1263 [hep-th]} \BibitemShut {NoStop}%
\bibitem [{\citenamefont {Perlmutter}(2014)}]{Perlmutter:2013gua}%
  \BibitemOpen
  \bibfield  {author} {\bibinfo {author} {\bibfnamefont {E.}~\bibnamefont
  {Perlmutter}},\ }\href {\doibase 10.1007/JHEP03(2014)117} {\bibfield
  {journal} {\bibinfo  {journal} {JHEP}\ }\textbf {\bibinfo {volume} {03}},\
  \bibinfo {pages} {117} (\bibinfo {year} {2014})},\ \Eprint
  {http://arxiv.org/abs/1308.1083} {arXiv:1308.1083 [hep-th]} \BibitemShut
  {NoStop}%
\bibitem [{\citenamefont {Mezei}(2015)}]{Mezei:2014zla}%
  \BibitemOpen
  \bibfield  {author} {\bibinfo {author} {\bibfnamefont {M.}~\bibnamefont
  {Mezei}},\ }\href {\doibase 10.1103/PhysRevD.91.045038} {\bibfield  {journal}
  {\bibinfo  {journal} {Phys. Rev.}\ }\textbf {\bibinfo {volume} {D91}},\
  \bibinfo {pages} {045038} (\bibinfo {year} {2015})},\ \Eprint
  {http://arxiv.org/abs/1411.7011} {arXiv:1411.7011 [hep-th]} \BibitemShut
  {NoStop}%
\bibitem [{\citenamefont {Bueno}\ \emph {et~al.}(2015)\citenamefont {Bueno},
  \citenamefont {Myers},\ and\ \citenamefont {Witczak-Krempa}}]{Bueno:2015rda}%
  \BibitemOpen
  \bibfield  {author} {\bibinfo {author} {\bibfnamefont {P.}~\bibnamefont
  {Bueno}}, \bibinfo {author} {\bibfnamefont {R.~C.}\ \bibnamefont {Myers}}, \
  and\ \bibinfo {author} {\bibfnamefont {W.}~\bibnamefont {Witczak-Krempa}},\
  }\href {\doibase 10.1103/PhysRevLett.115.021602} {\bibfield  {journal}
  {\bibinfo  {journal} {Phys. Rev. Lett.}\ }\textbf {\bibinfo {volume} {115}},\
  \bibinfo {pages} {021602} (\bibinfo {year} {2015})},\ \Eprint
  {http://arxiv.org/abs/1505.04804} {arXiv:1505.04804 [hep-th]} \BibitemShut
  {NoStop}%
\bibitem [{\citenamefont {Chu}\ and\ \citenamefont {Miao}(2016)}]{Chu:2016tps}%
  \BibitemOpen
  \bibfield  {author} {\bibinfo {author} {\bibfnamefont {C.-S.}\ \bibnamefont
  {Chu}}\ and\ \bibinfo {author} {\bibfnamefont {R.-X.}\ \bibnamefont {Miao}},\
  }\href {\doibase 10.1007/JHEP12(2016)036} {\bibfield  {journal} {\bibinfo
  {journal} {JHEP}\ }\textbf {\bibinfo {volume} {12}},\ \bibinfo {pages} {036}
  (\bibinfo {year} {2016})},\ \Eprint {http://arxiv.org/abs/1608.00328}
  {arXiv:1608.00328 [hep-th]} \BibitemShut {NoStop}%
\bibitem [{\citenamefont {Li}\ \emph {et~al.}(2018)\citenamefont {Li},
  \citenamefont {Lü},\ and\ \citenamefont {Mai}}]{Li:2018drw}%
  \BibitemOpen
  \bibfield  {author} {\bibinfo {author} {\bibfnamefont {Y.-Z.}\ \bibnamefont
  {Li}}, \bibinfo {author} {\bibfnamefont {H.}~\bibnamefont {Lü}}, \ and\
  \bibinfo {author} {\bibfnamefont {Z.-F.}\ \bibnamefont {Mai}},\ }\href
  {\doibase 10.1007/JHEP10(2018)063} {\bibfield  {journal} {\bibinfo  {journal}
  {JHEP}\ }\textbf {\bibinfo {volume} {10}},\ \bibinfo {pages} {063} (\bibinfo
  {year} {2018})},\ \Eprint {http://arxiv.org/abs/1808.00494} {arXiv:1808.00494
  [hep-th]} \BibitemShut {NoStop}%
\bibitem [{\citenamefont {Bueno}\ \emph
  {et~al.}(2019{\natexlab{a}})\citenamefont {Bueno}, \citenamefont {Cano},
  \citenamefont {Hennigar},\ and\ \citenamefont {Mann}}]{Bueno:2018yzo}%
  \BibitemOpen
  \bibfield  {author} {\bibinfo {author} {\bibfnamefont {P.}~\bibnamefont
  {Bueno}}, \bibinfo {author} {\bibfnamefont {P.~A.}\ \bibnamefont {Cano}},
  \bibinfo {author} {\bibfnamefont {R.~A.}\ \bibnamefont {Hennigar}}, \ and\
  \bibinfo {author} {\bibfnamefont {R.~B.}\ \bibnamefont {Mann}},\ }\href
  {\doibase 10.1103/PhysRevLett.122.071602} {\bibfield  {journal} {\bibinfo
  {journal} {Phys. Rev. Lett.}\ }\textbf {\bibinfo {volume} {122}},\ \bibinfo
  {pages} {071602} (\bibinfo {year} {2019}{\natexlab{a}})},\ \Eprint
  {http://arxiv.org/abs/1808.02052} {arXiv:1808.02052 [hep-th]} \BibitemShut
  {NoStop}%
\bibitem [{\citenamefont {Bueno}\ \emph {et~al.}(2022)\citenamefont {Bueno},
  \citenamefont {Cano}, \citenamefont {Murcia},\ and\ \citenamefont
  {Rivadulla~S\'anchez}}]{Bueno:2022jbl}%
  \BibitemOpen
  \bibfield  {author} {\bibinfo {author} {\bibfnamefont {P.}~\bibnamefont
  {Bueno}}, \bibinfo {author} {\bibfnamefont {P.~A.}\ \bibnamefont {Cano}},
  \bibinfo {author} {\bibfnamefont {{\'A}.}~\bibnamefont {Murcia}}, \ and\
  \bibinfo {author} {\bibfnamefont {A.}~\bibnamefont {Rivadulla~S\'anchez}},\
  }\href {\doibase 10.1103/PhysRevLett.129.021601} {\bibfield  {journal}
  {\bibinfo  {journal} {Phys. Rev. Lett.}\ }\textbf {\bibinfo {volume} {129}},\
  \bibinfo {pages} {021601} (\bibinfo {year} {2022})},\ \Eprint
  {http://arxiv.org/abs/2203.04325} {arXiv:2203.04325 [hep-th]} \BibitemShut
  {NoStop}%
\bibitem [{\citenamefont {Baiguera}\ \emph {et~al.}(2022)\citenamefont
  {Baiguera}, \citenamefont {Bianchi}, \citenamefont {Chapman},\ and\
  \citenamefont {Galante}}]{Baiguera:2022sao}%
  \BibitemOpen
  \bibfield  {author} {\bibinfo {author} {\bibfnamefont {S.}~\bibnamefont
  {Baiguera}}, \bibinfo {author} {\bibfnamefont {L.}~\bibnamefont {Bianchi}},
  \bibinfo {author} {\bibfnamefont {S.}~\bibnamefont {Chapman}}, \ and\
  \bibinfo {author} {\bibfnamefont {D.~A.}\ \bibnamefont {Galante}},\ }\href
  {\doibase 10.1007/JHEP06(2022)068} {\bibfield  {journal} {\bibinfo  {journal}
  {JHEP}\ }\textbf {\bibinfo {volume} {06}},\ \bibinfo {pages} {068} (\bibinfo
  {year} {2022})},\ \Eprint {http://arxiv.org/abs/2203.15028} {arXiv:2203.15028
  [hep-th]} \BibitemShut {NoStop}%
\bibitem [{\citenamefont {Kats}\ and\ \citenamefont
  {Petrov}(2009)}]{Kats:2007mq}%
  \BibitemOpen
  \bibfield  {author} {\bibinfo {author} {\bibfnamefont {Y.}~\bibnamefont
  {Kats}}\ and\ \bibinfo {author} {\bibfnamefont {P.}~\bibnamefont {Petrov}},\
  }\href {\doibase 10.1088/1126-6708/2009/01/044} {\bibfield  {journal}
  {\bibinfo  {journal} {JHEP}\ }\textbf {\bibinfo {volume} {01}},\ \bibinfo
  {pages} {044} (\bibinfo {year} {2009})},\ \Eprint
  {http://arxiv.org/abs/0712.0743} {arXiv:0712.0743 [hep-th]} \BibitemShut
  {NoStop}%
\bibitem [{\citenamefont {Brigante}\ \emph
  {et~al.}(2008{\natexlab{a}})\citenamefont {Brigante}, \citenamefont {Liu},
  \citenamefont {Myers}, \citenamefont {Shenker},\ and\ \citenamefont
  {Yaida}}]{Brigante:2007nu}%
  \BibitemOpen
  \bibfield  {author} {\bibinfo {author} {\bibfnamefont {M.}~\bibnamefont
  {Brigante}}, \bibinfo {author} {\bibfnamefont {H.}~\bibnamefont {Liu}},
  \bibinfo {author} {\bibfnamefont {R.~C.}\ \bibnamefont {Myers}}, \bibinfo
  {author} {\bibfnamefont {S.}~\bibnamefont {Shenker}}, \ and\ \bibinfo
  {author} {\bibfnamefont {S.}~\bibnamefont {Yaida}},\ }\href {\doibase
  10.1103/PhysRevD.77.126006} {\bibfield  {journal} {\bibinfo  {journal} {Phys.
  Rev.}\ }\textbf {\bibinfo {volume} {D77}},\ \bibinfo {pages} {126006}
  (\bibinfo {year} {2008}{\natexlab{a}})},\ \Eprint
  {http://arxiv.org/abs/0712.0805} {arXiv:0712.0805 [hep-th]} \BibitemShut
  {NoStop}%
\bibitem [{\citenamefont {Ge}\ \emph {et~al.}(2008)\citenamefont {Ge},
  \citenamefont {Matsuo}, \citenamefont {Shu}, \citenamefont {Sin},\ and\
  \citenamefont {Tsukioka}}]{Ge:2008ni}%
  \BibitemOpen
  \bibfield  {author} {\bibinfo {author} {\bibfnamefont {X.-H.}\ \bibnamefont
  {Ge}}, \bibinfo {author} {\bibfnamefont {Y.}~\bibnamefont {Matsuo}}, \bibinfo
  {author} {\bibfnamefont {F.-W.}\ \bibnamefont {Shu}}, \bibinfo {author}
  {\bibfnamefont {S.-J.}\ \bibnamefont {Sin}}, \ and\ \bibinfo {author}
  {\bibfnamefont {T.}~\bibnamefont {Tsukioka}},\ }\href {\doibase
  10.1088/1126-6708/2008/10/009} {\bibfield  {journal} {\bibinfo  {journal}
  {JHEP}\ }\textbf {\bibinfo {volume} {10}},\ \bibinfo {pages} {009} (\bibinfo
  {year} {2008})},\ \Eprint {http://arxiv.org/abs/0808.2354} {arXiv:0808.2354
  [hep-th]} \BibitemShut {NoStop}%
\bibitem [{\citenamefont {Gregory}\ \emph {et~al.}(2009)\citenamefont
  {Gregory}, \citenamefont {Kanno},\ and\ \citenamefont
  {Soda}}]{Gregory:2009fj}%
  \BibitemOpen
  \bibfield  {author} {\bibinfo {author} {\bibfnamefont {R.}~\bibnamefont
  {Gregory}}, \bibinfo {author} {\bibfnamefont {S.}~\bibnamefont {Kanno}}, \
  and\ \bibinfo {author} {\bibfnamefont {J.}~\bibnamefont {Soda}},\ }\href
  {\doibase 10.1088/1126-6708/2009/10/010} {\bibfield  {journal} {\bibinfo
  {journal} {JHEP}\ }\textbf {\bibinfo {volume} {10}},\ \bibinfo {pages} {010}
  (\bibinfo {year} {2009})},\ \Eprint {http://arxiv.org/abs/0907.3203}
  {arXiv:0907.3203 [hep-th]} \BibitemShut {NoStop}%
\bibitem [{\citenamefont {Pan}\ and\ \citenamefont {Wang}(2010)}]{Pan:2010at}%
  \BibitemOpen
  \bibfield  {author} {\bibinfo {author} {\bibfnamefont {Q.}~\bibnamefont
  {Pan}}\ and\ \bibinfo {author} {\bibfnamefont {B.}~\bibnamefont {Wang}},\
  }\href {\doibase 10.1016/j.physletb.2010.08.017} {\bibfield  {journal}
  {\bibinfo  {journal} {Phys. Lett. B}\ }\textbf {\bibinfo {volume} {693}},\
  \bibinfo {pages} {159} (\bibinfo {year} {2010})},\ \Eprint
  {http://arxiv.org/abs/1005.4743} {arXiv:1005.4743 [hep-th]} \BibitemShut
  {NoStop}%
\bibitem [{\citenamefont {Edelstein}\ \emph {et~al.}(2022)\citenamefont
  {Edelstein}, \citenamefont {Grandi},\ and\ \citenamefont
  {Rivadulla~S\'anchez}}]{Edelstein:2022xlb}%
  \BibitemOpen
  \bibfield  {author} {\bibinfo {author} {\bibfnamefont {J.~D.}\ \bibnamefont
  {Edelstein}}, \bibinfo {author} {\bibfnamefont {N.}~\bibnamefont {Grandi}}, \
  and\ \bibinfo {author} {\bibfnamefont {A.}~\bibnamefont
  {Rivadulla~S\'anchez}},\ }\href {\doibase 10.1007/JHEP05(2022)188} {\bibfield
   {journal} {\bibinfo  {journal} {JHEP}\ }\textbf {\bibinfo {volume} {05}},\
  \bibinfo {pages} {188} (\bibinfo {year} {2022})},\ \Eprint
  {http://arxiv.org/abs/2202.05781} {arXiv:2202.05781 [hep-th]} \BibitemShut
  {NoStop}%
\bibitem [{\citenamefont {Myers}\ \emph {et~al.}(2011)\citenamefont {Myers},
  \citenamefont {Sachdev},\ and\ \citenamefont {Singh}}]{Myers:2010pk}%
  \BibitemOpen
  \bibfield  {author} {\bibinfo {author} {\bibfnamefont {R.~C.}\ \bibnamefont
  {Myers}}, \bibinfo {author} {\bibfnamefont {S.}~\bibnamefont {Sachdev}}, \
  and\ \bibinfo {author} {\bibfnamefont {A.}~\bibnamefont {Singh}},\ }\href
  {\doibase 10.1103/PhysRevD.83.066017} {\bibfield  {journal} {\bibinfo
  {journal} {Phys. Rev. D}\ }\textbf {\bibinfo {volume} {83}},\ \bibinfo
  {pages} {066017} (\bibinfo {year} {2011})},\ \Eprint
  {http://arxiv.org/abs/1010.0443} {arXiv:1010.0443 [hep-th]} \BibitemShut
  {NoStop}%
\bibitem [{\citenamefont {Pal}(2011)}]{Pal:2010sx}%
  \BibitemOpen
  \bibfield  {author} {\bibinfo {author} {\bibfnamefont {S.~S.}\ \bibnamefont
  {Pal}},\ }\href {\doibase 10.1103/PhysRevD.84.126009} {\bibfield  {journal}
  {\bibinfo  {journal} {Phys. Rev. D}\ }\textbf {\bibinfo {volume} {84}},\
  \bibinfo {pages} {126009} (\bibinfo {year} {2011})},\ \Eprint
  {http://arxiv.org/abs/1011.3117} {arXiv:1011.3117 [hep-th]} \BibitemShut
  {NoStop}%
\bibitem [{\citenamefont {Witczak-Krempa}\ and\ \citenamefont
  {Sachdev}(2012)}]{Witczak-Krempa:2012qgh}%
  \BibitemOpen
  \bibfield  {author} {\bibinfo {author} {\bibfnamefont {W.}~\bibnamefont
  {Witczak-Krempa}}\ and\ \bibinfo {author} {\bibfnamefont {S.}~\bibnamefont
  {Sachdev}},\ }\href {\doibase 10.1103/PhysRevB.86.235115} {\bibfield
  {journal} {\bibinfo  {journal} {Phys. Rev. B}\ }\textbf {\bibinfo {volume}
  {86}},\ \bibinfo {pages} {235115} (\bibinfo {year} {2012})},\ \Eprint
  {http://arxiv.org/abs/1210.4166} {arXiv:1210.4166 [cond-mat.str-el]}
  \BibitemShut {NoStop}%
\bibitem [{\citenamefont {Witczak-Krempa}(2014)}]{Witczak-Krempa:2013aea}%
  \BibitemOpen
  \bibfield  {author} {\bibinfo {author} {\bibfnamefont {W.}~\bibnamefont
  {Witczak-Krempa}},\ }\href {\doibase 10.1103/PhysRevB.89.161114} {\bibfield
  {journal} {\bibinfo  {journal} {Phys. Rev. B}\ }\textbf {\bibinfo {volume}
  {89}},\ \bibinfo {pages} {161114} (\bibinfo {year} {2014})},\ \Eprint
  {http://arxiv.org/abs/1312.3334} {arXiv:1312.3334 [cond-mat.str-el]}
  \BibitemShut {NoStop}%
\bibitem [{\citenamefont {Ling}\ \emph {et~al.}(2017)\citenamefont {Ling},
  \citenamefont {Liu}, \citenamefont {Wu},\ and\ \citenamefont
  {Zhou}}]{Ling:2016dck}%
  \BibitemOpen
  \bibfield  {author} {\bibinfo {author} {\bibfnamefont {Y.}~\bibnamefont
  {Ling}}, \bibinfo {author} {\bibfnamefont {P.}~\bibnamefont {Liu}}, \bibinfo
  {author} {\bibfnamefont {J.-P.}\ \bibnamefont {Wu}}, \ and\ \bibinfo {author}
  {\bibfnamefont {Z.}~\bibnamefont {Zhou}},\ }\href {\doibase
  10.1016/j.physletb.2016.12.051} {\bibfield  {journal} {\bibinfo  {journal}
  {Phys. Lett. B}\ }\textbf {\bibinfo {volume} {766}},\ \bibinfo {pages} {41}
  (\bibinfo {year} {2017})},\ \Eprint {http://arxiv.org/abs/1606.07866}
  {arXiv:1606.07866 [hep-th]} \BibitemShut {NoStop}%
\bibitem [{\citenamefont {Sorokin}(2022)}]{Sorokin:2021tge}%
  \BibitemOpen
  \bibfield  {author} {\bibinfo {author} {\bibfnamefont {D.~P.}\ \bibnamefont
  {Sorokin}},\ }\href {\doibase 10.1002/prop.202200092} {\bibfield  {journal}
  {\bibinfo  {journal} {Fortsch. Phys.}\ }\textbf {\bibinfo {volume} {70}},\
  \bibinfo {pages} {2200092} (\bibinfo {year} {2022})},\ \Eprint
  {http://arxiv.org/abs/2112.12118} {arXiv:2112.12118 [hep-th]} \BibitemShut
  {NoStop}%
\bibitem [{\citenamefont {Cano}\ and\ \citenamefont
  {Murcia}(2021{\natexlab{a}})}]{Cano:2021tfs}%
  \BibitemOpen
  \bibfield  {author} {\bibinfo {author} {\bibfnamefont {P.~A.}\ \bibnamefont
  {Cano}}\ and\ \bibinfo {author} {\bibfnamefont {{\'A}.}~\bibnamefont
  {Murcia}},\ }\href {\doibase 10.1007/JHEP08(2021)042} {\bibfield  {journal}
  {\bibinfo  {journal} {JHEP}\ }\textbf {\bibinfo {volume} {08}},\ \bibinfo
  {pages} {042} (\bibinfo {year} {2021}{\natexlab{a}})},\ \Eprint
  {http://arxiv.org/abs/2104.07674} {arXiv:2104.07674 [hep-th]} \BibitemShut
  {NoStop}%
\bibitem [{\citenamefont {Cano}\ and\ \citenamefont
  {Murcia}(2021{\natexlab{b}})}]{Cano:2021hje}%
  \BibitemOpen
  \bibfield  {author} {\bibinfo {author} {\bibfnamefont {P.~A.}\ \bibnamefont
  {Cano}}\ and\ \bibinfo {author} {\bibfnamefont {{\'A}.}~\bibnamefont
  {Murcia}},\ }\href {\doibase 10.1103/PhysRevD.104.L101501} {\bibfield
  {journal} {\bibinfo  {journal} {Phys. Rev. D}\ }\textbf {\bibinfo {volume}
  {104}},\ \bibinfo {pages} {L101501} (\bibinfo {year} {2021}{\natexlab{b}})},\
  \Eprint {http://arxiv.org/abs/2105.09868} {arXiv:2105.09868 [hep-th]}
  \BibitemShut {NoStop}%
\bibitem [{\citenamefont {Weinberg}(2005)}]{Weinberg:1995mt}%
  \BibitemOpen
  \bibfield  {author} {\bibinfo {author} {\bibfnamefont {S.}~\bibnamefont
  {Weinberg}},\ }\href {\doibase 10.1017/CBO9781139644167} {\emph {\bibinfo
  {title} {{The Quantum theory of fields. Vol. 1: Foundations}}}}\ (\bibinfo
  {publisher} {Cambridge University Press},\ \bibinfo {year}
  {2005})\BibitemShut {NoStop}%
\bibitem [{Note1()}]{Note1}%
  \BibitemOpen
  \bibinfo {note} {Observe that $\protect \tensor {\chi }{_{\mu \nu }^{\rho
  \sigma }}$ can be compactly expressed as $\protect \tensor {\chi }{_{\mu \nu
  }^{\rho \sigma }}=\protect \tensor {\Theta }{_{\mu \nu }^{\rho \sigma
  }}+\protect \sqrt {\protect \tensor {\delta }{_{[\mu }^{[\rho }} \protect
  \tensor {\delta }{_{\nu ]}^{\sigma ]}}+\protect \tensor {\Theta }{^{2}_{\mu
  \nu }^{\rho \sigma }}}$, as in \cite {Cano:2021hje}.}\BibitemShut {Stop}%
\bibitem [{\citenamefont {Bueno}\ and\ \citenamefont
  {Cano}(2016)}]{Bueno:2016xff}%
  \BibitemOpen
  \bibfield  {author} {\bibinfo {author} {\bibfnamefont {P.}~\bibnamefont
  {Bueno}}\ and\ \bibinfo {author} {\bibfnamefont {P.~A.}\ \bibnamefont
  {Cano}},\ }\href {\doibase 10.1103/PhysRevD.94.104005} {\bibfield  {journal}
  {\bibinfo  {journal} {Phys. Rev.}\ }\textbf {\bibinfo {volume} {D94}},\
  \bibinfo {pages} {104005} (\bibinfo {year} {2016})},\ \Eprint
  {http://arxiv.org/abs/1607.06463} {arXiv:1607.06463 [hep-th]} \BibitemShut
  {NoStop}%
\bibitem [{\citenamefont {Bueno}\ \emph {et~al.}(2017)\citenamefont {Bueno},
  \citenamefont {Cano}, \citenamefont {Min},\ and\ \citenamefont
  {Visser}}]{Bueno:2016ypa}%
  \BibitemOpen
  \bibfield  {author} {\bibinfo {author} {\bibfnamefont {P.}~\bibnamefont
  {Bueno}}, \bibinfo {author} {\bibfnamefont {P.~A.}\ \bibnamefont {Cano}},
  \bibinfo {author} {\bibfnamefont {V.~S.}\ \bibnamefont {Min}}, \ and\
  \bibinfo {author} {\bibfnamefont {M.~R.}\ \bibnamefont {Visser}},\ }\href
  {\doibase 10.1103/PhysRevD.95.044010} {\bibfield  {journal} {\bibinfo
  {journal} {Phys. Rev.}\ }\textbf {\bibinfo {volume} {D95}},\ \bibinfo {pages}
  {044010} (\bibinfo {year} {2017})},\ \Eprint
  {http://arxiv.org/abs/1610.08519} {arXiv:1610.08519 [hep-th]} \BibitemShut
  {NoStop}%
\bibitem [{\citenamefont {Herzog}\ \emph {et~al.}(2007)\citenamefont {Herzog},
  \citenamefont {Kovtun}, \citenamefont {Sachdev},\ and\ \citenamefont
  {Son}}]{Herzog:2007ij}%
  \BibitemOpen
  \bibfield  {author} {\bibinfo {author} {\bibfnamefont {C.~P.}\ \bibnamefont
  {Herzog}}, \bibinfo {author} {\bibfnamefont {P.}~\bibnamefont {Kovtun}},
  \bibinfo {author} {\bibfnamefont {S.}~\bibnamefont {Sachdev}}, \ and\
  \bibinfo {author} {\bibfnamefont {D.~T.}\ \bibnamefont {Son}},\ }\href
  {\doibase 10.1103/PhysRevD.75.085020} {\bibfield  {journal} {\bibinfo
  {journal} {Phys. Rev. D}\ }\textbf {\bibinfo {volume} {75}},\ \bibinfo
  {pages} {085020} (\bibinfo {year} {2007})},\ \Eprint
  {http://arxiv.org/abs/hep-th/0701036} {arXiv:hep-th/0701036} \BibitemShut
  {NoStop}%
\bibitem [{Note2()}]{Note2}%
  \BibitemOpen
  \bibinfo {note} {Expressed in this way, it might appear that the equations of
  motion are of third order in derivatives. However, they can be seen to be
  equivalent to the (second-order) equations $\nabla _\mu \left ( \chi ^{\mu
  \nu \rho \sigma } F_{\rho \sigma } \right )=0$ in the gauge $A_u=0$ by
  appropriate manipulations which allow to integrate them into second-order
  equations, uniquely fixed after requiring $A_\mu =0$ to be a
  solution.}\BibitemShut {Stop}%
\bibitem [{\citenamefont {Son}\ and\ \citenamefont
  {Starinets}(2002)}]{Son:2002sd}%
  \BibitemOpen
  \bibfield  {author} {\bibinfo {author} {\bibfnamefont {D.~T.}\ \bibnamefont
  {Son}}\ and\ \bibinfo {author} {\bibfnamefont {A.~O.}\ \bibnamefont
  {Starinets}},\ }\href {\doibase 10.1088/1126-6708/2002/09/042} {\bibfield
  {journal} {\bibinfo  {journal} {JHEP}\ }\textbf {\bibinfo {volume} {09}},\
  \bibinfo {pages} {042} (\bibinfo {year} {2002})},\ \Eprint
  {http://arxiv.org/abs/hep-th/0205051} {arXiv:hep-th/0205051} \BibitemShut
  {NoStop}%
\bibitem [{\citenamefont {Policastro}\ \emph {et~al.}(2002)\citenamefont
  {Policastro}, \citenamefont {Son},\ and\ \citenamefont
  {Starinets}}]{Policastro:2002se}%
  \BibitemOpen
  \bibfield  {author} {\bibinfo {author} {\bibfnamefont {G.}~\bibnamefont
  {Policastro}}, \bibinfo {author} {\bibfnamefont {D.~T.}\ \bibnamefont {Son}},
  \ and\ \bibinfo {author} {\bibfnamefont {A.~O.}\ \bibnamefont {Starinets}},\
  }\href {\doibase 10.1088/1126-6708/2002/09/043} {\bibfield  {journal}
  {\bibinfo  {journal} {JHEP}\ }\textbf {\bibinfo {volume} {09}},\ \bibinfo
  {pages} {043} (\bibinfo {year} {2002})},\ \Eprint
  {http://arxiv.org/abs/hep-th/0205052} {arXiv:hep-th/0205052} \BibitemShut
  {NoStop}%
\bibitem [{Note3()}]{Note3}%
  \BibitemOpen
  \bibinfo {note} {Boundary indices are lowered and raised with the Minkowski
  metric $\eta ^{ab}$.}\BibitemShut {Stop}%
\bibitem [{Note4()}]{Note4}%
  \BibitemOpen
  \bibinfo {note} {Up to a constant on the right-hand side of \protect \textup
  {\hbox {\mathsurround \z@ \protect \normalfont (\ignorespaces \ref
  {eq:univresse}\unskip \@@italiccorr )}}, which corresponds to the different
  conventions used for the electromagnetic duality transformation.}\BibitemShut
  {Stop}%
\bibitem [{\citenamefont {Kubo}(1957)}]{kubo1957statistical}%
  \BibitemOpen
  \bibfield  {author} {\bibinfo {author} {\bibfnamefont {R.}~\bibnamefont
  {Kubo}},\ }\href@noop {} {\bibfield  {journal} {\bibinfo  {journal} {Journal
  of the Physical Society of Japan}\ }\textbf {\bibinfo {volume} {12}},\
  \bibinfo {pages} {570} (\bibinfo {year} {1957})}\BibitemShut {NoStop}%
\bibitem [{Note5()}]{Note5}%
  \BibitemOpen
  \bibinfo {note} {After choosing the sign for $K^L(\omega ,0)=K^T(\omega ,0)$
  that guarantees a positive spectral function, defined by $-2\protect \,
  \protect \mathrm {Im}(\protect \sqrt {k^2-\omega ^2} K^T)$. The same result
  could also be derived by direct resolution of \protect \textup {\hbox
  {\mathsurround \z@ \protect \normalfont (\ignorespaces \ref {eq:symA}\unskip
  \@@italiccorr )}} for $k=0$.}\BibitemShut {Stop}%
\bibitem [{Note6()}]{Note6}%
  \BibitemOpen
  \bibinfo {note} {This hydrodynamic-to-collisionless crossover manifests in
  the longitudinal conductivity and not in the transverse one since the
  longitudinal correlator $C_{xx}$ is related via Eq. \protect \textup {\hbox
  {\mathsurround \z@ \protect \normalfont (\ignorespaces \ref {eq:clong}\unskip
  \@@italiccorr )}} to the density-density correlator $C_{tt}$, which is a
  clear probe of hydrodynamic behaviour \cite {Herzog:2007ij}.}\BibitemShut
  {Stop}%
\bibitem [{\citenamefont {Kovtun}\ \emph {et~al.}(2003)\citenamefont {Kovtun},
  \citenamefont {Son},\ and\ \citenamefont {Starinets}}]{Kovtun:2003wp}%
  \BibitemOpen
  \bibfield  {author} {\bibinfo {author} {\bibfnamefont {P.}~\bibnamefont
  {Kovtun}}, \bibinfo {author} {\bibfnamefont {D.~T.}\ \bibnamefont {Son}}, \
  and\ \bibinfo {author} {\bibfnamefont {A.~O.}\ \bibnamefont {Starinets}},\
  }\href {\doibase 10.1088/1126-6708/2003/10/064} {\bibfield  {journal}
  {\bibinfo  {journal} {JHEP}\ }\textbf {\bibinfo {volume} {10}},\ \bibinfo
  {pages} {064} (\bibinfo {year} {2003})},\ \Eprint
  {http://arxiv.org/abs/hep-th/0309213} {arXiv:hep-th/0309213} \BibitemShut
  {NoStop}%
\bibitem [{\citenamefont {Iqbal}\ and\ \citenamefont
  {Liu}(2009)}]{Iqbal:2008by}%
  \BibitemOpen
  \bibfield  {author} {\bibinfo {author} {\bibfnamefont {N.}~\bibnamefont
  {Iqbal}}\ and\ \bibinfo {author} {\bibfnamefont {H.}~\bibnamefont {Liu}},\
  }\href {\doibase 10.1103/PhysRevD.79.025023} {\bibfield  {journal} {\bibinfo
  {journal} {Phys. Rev. D}\ }\textbf {\bibinfo {volume} {79}},\ \bibinfo
  {pages} {025023} (\bibinfo {year} {2009})},\ \Eprint
  {http://arxiv.org/abs/0809.3808} {arXiv:0809.3808 [hep-th]} \BibitemShut
  {NoStop}%
\bibitem [{\citenamefont {Buchdahl}(1970)}]{Buchdahl:1970ynr}%
  \BibitemOpen
  \bibfield  {author} {\bibinfo {author} {\bibfnamefont {H.~A.}\ \bibnamefont
  {Buchdahl}},\ }\href@noop {} {\bibfield  {journal} {\bibinfo  {journal} {Mon.
  Not. Roy. Astron. Soc.}\ }\textbf {\bibinfo {volume} {150}},\ \bibinfo
  {pages} {1} (\bibinfo {year} {1970})}\BibitemShut {NoStop}%
\bibitem [{\citenamefont {Sotiriou}\ and\ \citenamefont
  {Faraoni}(2010)}]{Sotiriou:2008rp}%
  \BibitemOpen
  \bibfield  {author} {\bibinfo {author} {\bibfnamefont {T.~P.}\ \bibnamefont
  {Sotiriou}}\ and\ \bibinfo {author} {\bibfnamefont {V.}~\bibnamefont
  {Faraoni}},\ }\href {\doibase 10.1103/RevModPhys.82.451} {\bibfield
  {journal} {\bibinfo  {journal} {Rev. Mod. Phys.}\ }\textbf {\bibinfo {volume}
  {82}},\ \bibinfo {pages} {451} (\bibinfo {year} {2010})},\ \Eprint
  {http://arxiv.org/abs/0805.1726} {arXiv:0805.1726 [gr-qc]} \BibitemShut
  {NoStop}%
\bibitem [{Note7()}]{Note7}%
  \BibitemOpen
  \bibinfo {note} {If one insists on working with the GR solution, the
  lowest-order choice for $\protect \mathcal {T}_{\mu \nu }$ which is nonzero
  on the AdS black brane solution of GR is the traceless part of $\nabla _\mu
  W_{\alpha \beta \rho \sigma } \nabla _\nu W^{\alpha \beta \rho \sigma }$,
  where $W_{\alpha \beta \rho \sigma }$ stands for the Weyl tensor. This term
  would in principle modify the conductivities of the dual CFT. Nevertheless,
  such a term is of order $\protect \mathcal {O}(L^8)$ and there are other
  higher-curvature terms of less order that already modify the gravitational
  background.}\BibitemShut {Stop}%
\bibitem [{\citenamefont {Arkani-Hamed}\ \emph {et~al.}(2007)\citenamefont
  {Arkani-Hamed}, \citenamefont {Motl}, \citenamefont {Nicolis},\ and\
  \citenamefont {Vafa}}]{Arkani-Hamed:2006emk}%
  \BibitemOpen
  \bibfield  {author} {\bibinfo {author} {\bibfnamefont {N.}~\bibnamefont
  {Arkani-Hamed}}, \bibinfo {author} {\bibfnamefont {L.}~\bibnamefont {Motl}},
  \bibinfo {author} {\bibfnamefont {A.}~\bibnamefont {Nicolis}}, \ and\
  \bibinfo {author} {\bibfnamefont {C.}~\bibnamefont {Vafa}},\ }\href {\doibase
  10.1088/1126-6708/2007/06/060} {\bibfield  {journal} {\bibinfo  {journal}
  {JHEP}\ }\textbf {\bibinfo {volume} {06}},\ \bibinfo {pages} {060} (\bibinfo
  {year} {2007})},\ \Eprint {http://arxiv.org/abs/hep-th/0601001}
  {arXiv:hep-th/0601001} \BibitemShut {NoStop}%
\bibitem [{\citenamefont {Bueno}\ \emph
  {et~al.}(2019{\natexlab{b}})\citenamefont {Bueno}, \citenamefont {Cano},
  \citenamefont {Moreno},\ and\ \citenamefont {Murcia}}]{Bueno:2019ltp}%
  \BibitemOpen
  \bibfield  {author} {\bibinfo {author} {\bibfnamefont {P.}~\bibnamefont
  {Bueno}}, \bibinfo {author} {\bibfnamefont {P.~A.}\ \bibnamefont {Cano}},
  \bibinfo {author} {\bibfnamefont {J.}~\bibnamefont {Moreno}}, \ and\ \bibinfo
  {author} {\bibfnamefont {{\'A}.}~\bibnamefont {Murcia}},\ }\href {\doibase
  10.1007/JHEP11(2019)062} {\bibfield  {journal} {\bibinfo  {journal} {JHEP}\
  }\textbf {\bibinfo {volume} {11}},\ \bibinfo {pages} {062} (\bibinfo {year}
  {2019}{\natexlab{b}})},\ \Eprint {http://arxiv.org/abs/1906.00987}
  {arXiv:1906.00987 [hep-th]} \BibitemShut {NoStop}%
\bibitem [{\citenamefont {Osborn}\ and\ \citenamefont
  {Petkou}(1994)}]{Osborn:1993cr}%
  \BibitemOpen
  \bibfield  {author} {\bibinfo {author} {\bibfnamefont {H.}~\bibnamefont
  {Osborn}}\ and\ \bibinfo {author} {\bibfnamefont {A.~C.}\ \bibnamefont
  {Petkou}},\ }\href {\doibase 10.1006/aphy.1994.1045} {\bibfield  {journal}
  {\bibinfo  {journal} {Annals Phys.}\ }\textbf {\bibinfo {volume} {231}},\
  \bibinfo {pages} {311} (\bibinfo {year} {1994})},\ \Eprint
  {http://arxiv.org/abs/hep-th/9307010} {arXiv:hep-th/9307010} \BibitemShut
  {NoStop}%
\bibitem [{\citenamefont {Erdmenger}\ and\ \citenamefont
  {Osborn}(1997)}]{Erdmenger:1996yc}%
  \BibitemOpen
  \bibfield  {author} {\bibinfo {author} {\bibfnamefont {J.}~\bibnamefont
  {Erdmenger}}\ and\ \bibinfo {author} {\bibfnamefont {H.}~\bibnamefont
  {Osborn}},\ }\href {\doibase 10.1016/S0550-3213(96)00545-7} {\bibfield
  {journal} {\bibinfo  {journal} {Nucl. Phys. B}\ }\textbf {\bibinfo {volume}
  {483}},\ \bibinfo {pages} {431} (\bibinfo {year} {1997})},\ \Eprint
  {http://arxiv.org/abs/hep-th/9605009} {arXiv:hep-th/9605009} \BibitemShut
  {NoStop}%
\bibitem [{\citenamefont {Hofman}\ and\ \citenamefont
  {Maldacena}(2008)}]{Hofman:2008ar}%
  \BibitemOpen
  \bibfield  {author} {\bibinfo {author} {\bibfnamefont {D.~M.}\ \bibnamefont
  {Hofman}}\ and\ \bibinfo {author} {\bibfnamefont {J.}~\bibnamefont
  {Maldacena}},\ }\href {\doibase 10.1088/1126-6708/2008/05/012} {\bibfield
  {journal} {\bibinfo  {journal} {JHEP}\ }\textbf {\bibinfo {volume} {05}},\
  \bibinfo {pages} {012} (\bibinfo {year} {2008})},\ \Eprint
  {http://arxiv.org/abs/0803.1467} {arXiv:0803.1467 [hep-th]} \BibitemShut
  {NoStop}%
\bibitem [{\citenamefont {Hartnoll}\ and\ \citenamefont
  {Kovtun}(2007)}]{Hartnoll:2007ai}%
  \BibitemOpen
  \bibfield  {author} {\bibinfo {author} {\bibfnamefont {S.~A.}\ \bibnamefont
  {Hartnoll}}\ and\ \bibinfo {author} {\bibfnamefont {P.}~\bibnamefont
  {Kovtun}},\ }\href {\doibase 10.1103/PhysRevD.76.066001} {\bibfield
  {journal} {\bibinfo  {journal} {Phys. Rev. D}\ }\textbf {\bibinfo {volume}
  {76}},\ \bibinfo {pages} {066001} (\bibinfo {year} {2007})},\ \Eprint
  {http://arxiv.org/abs/0704.1160} {arXiv:0704.1160 [hep-th]} \BibitemShut
  {NoStop}%
\bibitem [{\citenamefont {Hartnoll}\ \emph {et~al.}(2007)\citenamefont
  {Hartnoll}, \citenamefont {Kovtun}, \citenamefont {Muller},\ and\
  \citenamefont {Sachdev}}]{Hartnoll:2007ih}%
  \BibitemOpen
  \bibfield  {author} {\bibinfo {author} {\bibfnamefont {S.~A.}\ \bibnamefont
  {Hartnoll}}, \bibinfo {author} {\bibfnamefont {P.~K.}\ \bibnamefont
  {Kovtun}}, \bibinfo {author} {\bibfnamefont {M.}~\bibnamefont {Muller}}, \
  and\ \bibinfo {author} {\bibfnamefont {S.}~\bibnamefont {Sachdev}},\ }\href
  {\doibase 10.1103/PhysRevB.76.144502} {\bibfield  {journal} {\bibinfo
  {journal} {Phys. Rev. B}\ }\textbf {\bibinfo {volume} {76}},\ \bibinfo
  {pages} {144502} (\bibinfo {year} {2007})},\ \Eprint
  {http://arxiv.org/abs/0706.3215} {arXiv:0706.3215 [cond-mat.str-el]}
  \BibitemShut {NoStop}%
\bibitem [{\citenamefont {Goldstein}\ \emph {et~al.}(2010)\citenamefont
  {Goldstein}, \citenamefont {Kachru}, \citenamefont {Prakash},\ and\
  \citenamefont {Trivedi}}]{Goldstein:2009cv}%
  \BibitemOpen
  \bibfield  {author} {\bibinfo {author} {\bibfnamefont {K.}~\bibnamefont
  {Goldstein}}, \bibinfo {author} {\bibfnamefont {S.}~\bibnamefont {Kachru}},
  \bibinfo {author} {\bibfnamefont {S.}~\bibnamefont {Prakash}}, \ and\
  \bibinfo {author} {\bibfnamefont {S.~P.}\ \bibnamefont {Trivedi}},\ }\href
  {\doibase 10.1007/JHEP08(2010)078} {\bibfield  {journal} {\bibinfo  {journal}
  {JHEP}\ }\textbf {\bibinfo {volume} {08}},\ \bibinfo {pages} {078} (\bibinfo
  {year} {2010})},\ \Eprint {http://arxiv.org/abs/0911.3586} {arXiv:0911.3586
  [hep-th]} \BibitemShut {NoStop}%
\bibitem [{\citenamefont {Blake}\ and\ \citenamefont
  {Donos}(2015)}]{Blake:2014yla}%
  \BibitemOpen
  \bibfield  {author} {\bibinfo {author} {\bibfnamefont {M.}~\bibnamefont
  {Blake}}\ and\ \bibinfo {author} {\bibfnamefont {A.}~\bibnamefont {Donos}},\
  }\href {\doibase 10.1103/PhysRevLett.114.021601} {\bibfield  {journal}
  {\bibinfo  {journal} {Phys. Rev. Lett.}\ }\textbf {\bibinfo {volume} {114}},\
  \bibinfo {pages} {021601} (\bibinfo {year} {2015})},\ \Eprint
  {http://arxiv.org/abs/1406.1659} {arXiv:1406.1659 [hep-th]} \BibitemShut
  {NoStop}%
\bibitem [{\citenamefont {Guo}\ \emph {et~al.}(2018)\citenamefont {Guo},
  \citenamefont {Wang},\ and\ \citenamefont {Yang}}]{Guo:2017bru}%
  \BibitemOpen
  \bibfield  {author} {\bibinfo {author} {\bibfnamefont {X.}~\bibnamefont
  {Guo}}, \bibinfo {author} {\bibfnamefont {P.}~\bibnamefont {Wang}}, \ and\
  \bibinfo {author} {\bibfnamefont {H.}~\bibnamefont {Yang}},\ }\href {\doibase
  10.1103/PhysRevD.98.026021} {\bibfield  {journal} {\bibinfo  {journal} {Phys.
  Rev. D}\ }\textbf {\bibinfo {volume} {98}},\ \bibinfo {pages} {026021}
  (\bibinfo {year} {2018})},\ \Eprint {http://arxiv.org/abs/1711.03298}
  {arXiv:1711.03298 [hep-th]} \BibitemShut {NoStop}%
\bibitem [{\citenamefont {Wang}\ \emph {et~al.}(2019)\citenamefont {Wang},
  \citenamefont {Wu},\ and\ \citenamefont {Yang}}]{Wang:2018hwg}%
  \BibitemOpen
  \bibfield  {author} {\bibinfo {author} {\bibfnamefont {P.}~\bibnamefont
  {Wang}}, \bibinfo {author} {\bibfnamefont {H.}~\bibnamefont {Wu}}, \ and\
  \bibinfo {author} {\bibfnamefont {H.}~\bibnamefont {Yang}},\ }\href {\doibase
  10.1140/epjc/s10052-018-6503-8} {\bibfield  {journal} {\bibinfo  {journal}
  {Eur. Phys. J. C}\ }\textbf {\bibinfo {volume} {79}},\ \bibinfo {pages} {6}
  (\bibinfo {year} {2019})},\ \Eprint {http://arxiv.org/abs/1805.07913}
  {arXiv:1805.07913 [hep-th]} \BibitemShut {NoStop}%
\bibitem [{\citenamefont {Kiczek}\ \emph {et~al.}(2021)\citenamefont {Kiczek},
  \citenamefont {Rogatko},\ and\ \citenamefont {Wysokinski}}]{Kiczek:2020ngg}%
  \BibitemOpen
  \bibfield  {author} {\bibinfo {author} {\bibfnamefont {B.}~\bibnamefont
  {Kiczek}}, \bibinfo {author} {\bibfnamefont {M.}~\bibnamefont {Rogatko}}, \
  and\ \bibinfo {author} {\bibfnamefont {K.~I.}\ \bibnamefont {Wysokinski}},\
  }\href {\doibase 10.1103/PhysRevD.104.086022} {\bibfield  {journal} {\bibinfo
   {journal} {Phys. Rev. D}\ }\textbf {\bibinfo {volume} {104}},\ \bibinfo
  {pages} {086022} (\bibinfo {year} {2021})},\ \Eprint
  {http://arxiv.org/abs/2010.13095} {arXiv:2010.13095 [hep-th]} \BibitemShut
  {NoStop}%
\bibitem [{\citenamefont {Hennigar}\ \emph {et~al.}(2017)\citenamefont
  {Hennigar}, \citenamefont {Kubiz\v{n}\'ak},\ and\ \citenamefont
  {Mann}}]{Hennigar:2017ego}%
  \BibitemOpen
  \bibfield  {author} {\bibinfo {author} {\bibfnamefont {R.~A.}\ \bibnamefont
  {Hennigar}}, \bibinfo {author} {\bibfnamefont {D.}~\bibnamefont
  {Kubiz\v{n}\'ak}}, \ and\ \bibinfo {author} {\bibfnamefont {R.~B.}\
  \bibnamefont {Mann}},\ }\href {\doibase 10.1103/PhysRevD.95.104042}
  {\bibfield  {journal} {\bibinfo  {journal} {Phys. Rev. D}\ }\textbf {\bibinfo
  {volume} {95}},\ \bibinfo {pages} {104042} (\bibinfo {year} {2017})},\
  \Eprint {http://arxiv.org/abs/1703.01631} {arXiv:1703.01631 [hep-th]}
  \BibitemShut {NoStop}%
\bibitem [{\citenamefont {Bueno}\ and\ \citenamefont
  {Cano}(2017)}]{Bueno:2017sui}%
  \BibitemOpen
  \bibfield  {author} {\bibinfo {author} {\bibfnamefont {P.}~\bibnamefont
  {Bueno}}\ and\ \bibinfo {author} {\bibfnamefont {P.~A.}\ \bibnamefont
  {Cano}},\ }\href {\doibase 10.1088/1361-6382/aa8056} {\bibfield  {journal}
  {\bibinfo  {journal} {Class. Quant. Grav.}\ }\textbf {\bibinfo {volume}
  {34}},\ \bibinfo {pages} {175008} (\bibinfo {year} {2017})},\ \Eprint
  {http://arxiv.org/abs/1703.04625} {arXiv:1703.04625 [hep-th]} \BibitemShut
  {NoStop}%
\bibitem [{\citenamefont {Feng}\ \emph {et~al.}(2017)\citenamefont {Feng},
  \citenamefont {Huang}, \citenamefont {Mai},\ and\ \citenamefont
  {Lu}}]{Feng:2017tev}%
  \BibitemOpen
  \bibfield  {author} {\bibinfo {author} {\bibfnamefont {X.-H.}\ \bibnamefont
  {Feng}}, \bibinfo {author} {\bibfnamefont {H.}~\bibnamefont {Huang}},
  \bibinfo {author} {\bibfnamefont {Z.-F.}\ \bibnamefont {Mai}}, \ and\
  \bibinfo {author} {\bibfnamefont {H.}~\bibnamefont {Lu}},\ }\href {\doibase
  10.1103/PhysRevD.96.104034} {\bibfield  {journal} {\bibinfo  {journal} {Phys.
  Rev.}\ }\textbf {\bibinfo {volume} {D96}},\ \bibinfo {pages} {104034}
  (\bibinfo {year} {2017})},\ \Eprint {http://arxiv.org/abs/1707.06308}
  {arXiv:1707.06308 [hep-th]} \BibitemShut {NoStop}%
\bibitem [{\citenamefont {Palais}(1979)}]{Palais:1979rca}%
  \BibitemOpen
  \bibfield  {author} {\bibinfo {author} {\bibfnamefont {R.~S.}\ \bibnamefont
  {Palais}},\ }\href {\doibase 10.1007/BF01941322} {\bibfield  {journal}
  {\bibinfo  {journal} {Commun. Math. Phys.}\ }\textbf {\bibinfo {volume}
  {69}},\ \bibinfo {pages} {19} (\bibinfo {year} {1979})}\BibitemShut {NoStop}%
\bibitem [{\citenamefont {Deser}\ and\ \citenamefont
  {Tekin}(2003)}]{Deser:2003up}%
  \BibitemOpen
  \bibfield  {author} {\bibinfo {author} {\bibfnamefont {S.}~\bibnamefont
  {Deser}}\ and\ \bibinfo {author} {\bibfnamefont {B.}~\bibnamefont {Tekin}},\
  }\href {\doibase 10.1088/0264-9381/20/22/011} {\bibfield  {journal} {\bibinfo
   {journal} {Class. Quant. Grav.}\ }\textbf {\bibinfo {volume} {20}},\
  \bibinfo {pages} {4877} (\bibinfo {year} {2003})},\ \Eprint
  {http://arxiv.org/abs/gr-qc/0306114} {arXiv:gr-qc/0306114} \BibitemShut
  {NoStop}%
\bibitem [{\citenamefont {Harlow}\ \emph {et~al.}(2022)\citenamefont {Harlow},
  \citenamefont {Heidenreich}, \citenamefont {Reece},\ and\ \citenamefont
  {Rudelius}}]{Harlow:2022gzl}%
  \BibitemOpen
  \bibfield  {author} {\bibinfo {author} {\bibfnamefont {D.}~\bibnamefont
  {Harlow}}, \bibinfo {author} {\bibfnamefont {B.}~\bibnamefont {Heidenreich}},
  \bibinfo {author} {\bibfnamefont {M.}~\bibnamefont {Reece}}, \ and\ \bibinfo
  {author} {\bibfnamefont {T.}~\bibnamefont {Rudelius}},\ }\href@noop {} {\
  (\bibinfo {year} {2022})},\ \Eprint {http://arxiv.org/abs/2201.08380}
  {arXiv:2201.08380 [hep-th]} \BibitemShut {NoStop}%
\bibitem [{\citenamefont {Cheung}\ \emph {et~al.}(2018)\citenamefont {Cheung},
  \citenamefont {Liu},\ and\ \citenamefont {Remmen}}]{Cheung:2018cwt}%
  \BibitemOpen
  \bibfield  {author} {\bibinfo {author} {\bibfnamefont {C.}~\bibnamefont
  {Cheung}}, \bibinfo {author} {\bibfnamefont {J.}~\bibnamefont {Liu}}, \ and\
  \bibinfo {author} {\bibfnamefont {G.~N.}\ \bibnamefont {Remmen}},\ }\href
  {\doibase 10.1007/JHEP10(2018)004} {\bibfield  {journal} {\bibinfo  {journal}
  {JHEP}\ }\textbf {\bibinfo {volume} {10}},\ \bibinfo {pages} {004} (\bibinfo
  {year} {2018})},\ \Eprint {http://arxiv.org/abs/1801.08546} {arXiv:1801.08546
  [hep-th]} \BibitemShut {NoStop}%
\bibitem [{\citenamefont {Hamada}\ \emph {et~al.}(2019)\citenamefont {Hamada},
  \citenamefont {Noumi},\ and\ \citenamefont {Shiu}}]{Hamada:2018dde}%
  \BibitemOpen
  \bibfield  {author} {\bibinfo {author} {\bibfnamefont {Y.}~\bibnamefont
  {Hamada}}, \bibinfo {author} {\bibfnamefont {T.}~\bibnamefont {Noumi}}, \
  and\ \bibinfo {author} {\bibfnamefont {G.}~\bibnamefont {Shiu}},\ }\href
  {\doibase 10.1103/PhysRevLett.123.051601} {\bibfield  {journal} {\bibinfo
  {journal} {Phys. Rev. Lett.}\ }\textbf {\bibinfo {volume} {123}},\ \bibinfo
  {pages} {051601} (\bibinfo {year} {2019})},\ \Eprint
  {http://arxiv.org/abs/1810.03637} {arXiv:1810.03637 [hep-th]} \BibitemShut
  {NoStop}%
\bibitem [{\citenamefont {Cremonini}\ \emph {et~al.}(2020)\citenamefont
  {Cremonini}, \citenamefont {Jones}, \citenamefont {Liu},\ and\ \citenamefont
  {McPeak}}]{Cremonini:2019wdk}%
  \BibitemOpen
  \bibfield  {author} {\bibinfo {author} {\bibfnamefont {S.}~\bibnamefont
  {Cremonini}}, \bibinfo {author} {\bibfnamefont {C.~R.~T.}\ \bibnamefont
  {Jones}}, \bibinfo {author} {\bibfnamefont {J.~T.}\ \bibnamefont {Liu}}, \
  and\ \bibinfo {author} {\bibfnamefont {B.}~\bibnamefont {McPeak}},\ }\href
  {\doibase 10.1007/JHEP09(2020)003} {\bibfield  {journal} {\bibinfo  {journal}
  {JHEP}\ }\textbf {\bibinfo {volume} {09}},\ \bibinfo {pages} {003} (\bibinfo
  {year} {2020})},\ \Eprint {http://arxiv.org/abs/1912.11161} {arXiv:1912.11161
  [hep-th]} \BibitemShut {NoStop}%
\bibitem [{\citenamefont {Goon}\ and\ \citenamefont
  {Penco}(2020)}]{Goon:2019faz}%
  \BibitemOpen
  \bibfield  {author} {\bibinfo {author} {\bibfnamefont {G.}~\bibnamefont
  {Goon}}\ and\ \bibinfo {author} {\bibfnamefont {R.}~\bibnamefont {Penco}},\
  }\href {\doibase 10.1103/PhysRevLett.124.101103} {\bibfield  {journal}
  {\bibinfo  {journal} {Phys. Rev. Lett.}\ }\textbf {\bibinfo {volume} {124}},\
  \bibinfo {pages} {101103} (\bibinfo {year} {2020})},\ \Eprint
  {http://arxiv.org/abs/1909.05254} {arXiv:1909.05254 [hep-th]} \BibitemShut
  {NoStop}%
\bibitem [{\citenamefont {McPeak}(2022)}]{McPeak:2021tvu}%
  \BibitemOpen
  \bibfield  {author} {\bibinfo {author} {\bibfnamefont {B.}~\bibnamefont
  {McPeak}},\ }\href {\doibase 10.1103/PhysRevD.105.L081901} {\bibfield
  {journal} {\bibinfo  {journal} {Phys. Rev. D}\ }\textbf {\bibinfo {volume}
  {105}},\ \bibinfo {pages} {L081901} (\bibinfo {year} {2022})},\ \Eprint
  {http://arxiv.org/abs/2112.13433} {arXiv:2112.13433 [hep-th]} \BibitemShut
  {NoStop}%
\bibitem [{Note8()}]{Note8}%
  \BibitemOpen
  \bibinfo {note} {The value $u_0=\protect \sqrt [3]{4}$ corresponds to the
  extremal black brane and, in such a case, we observe that $u_1$ diverges.
  Therefore, the expansion in $\lambda $ is not reliable if we are arbitrarily
  close to extremality. Nevertheless, a near-extremal perturbative study of the
  black brane may be carried out if one remains in a regime for which $\vert
  \protect \sqrt [3]{4}-u_0\vert >> \lambda $ (so that $\lambda u_1$ is still
  small compared to $u_0$).}\BibitemShut {Stop}%
\bibitem [{Note9()}]{Note9}%
  \BibitemOpen
  \bibinfo {note} {This constraint was obtained perturbatively and could not be
  reliable for large $\vert \lambda \vert $ (which would also imply the failure
  of the expansion \protect \textup {\hbox {\mathsurround \z@ \protect
  \normalfont (\ignorespaces \ref {eq:appexpan}\unskip \@@italiccorr )}}).
  Nevertheless, since causality will further restrict $\vert \lambda \vert $ to
  be less than $\sim 1/2$, we will not worry about further
  refinement.}\BibitemShut {Stop}%
\bibitem [{\citenamefont {Brigante}\ \emph
  {et~al.}(2008{\natexlab{b}})\citenamefont {Brigante}, \citenamefont {Liu},
  \citenamefont {Myers}, \citenamefont {Shenker},\ and\ \citenamefont
  {Yaida}}]{Brigante:2008gz}%
  \BibitemOpen
  \bibfield  {author} {\bibinfo {author} {\bibfnamefont {M.}~\bibnamefont
  {Brigante}}, \bibinfo {author} {\bibfnamefont {H.}~\bibnamefont {Liu}},
  \bibinfo {author} {\bibfnamefont {R.~C.}\ \bibnamefont {Myers}}, \bibinfo
  {author} {\bibfnamefont {S.}~\bibnamefont {Shenker}}, \ and\ \bibinfo
  {author} {\bibfnamefont {S.}~\bibnamefont {Yaida}},\ }\href {\doibase
  10.1103/PhysRevLett.100.191601} {\bibfield  {journal} {\bibinfo  {journal}
  {Phys. Rev. Lett.}\ }\textbf {\bibinfo {volume} {100}},\ \bibinfo {pages}
  {191601} (\bibinfo {year} {2008}{\natexlab{b}})},\ \Eprint
  {http://arxiv.org/abs/0802.3318} {arXiv:0802.3318 [hep-th]} \BibitemShut
  {NoStop}%
\bibitem [{\citenamefont {Buchel}\ and\ \citenamefont
  {Myers}(2009)}]{Buchel:2009tt}%
  \BibitemOpen
  \bibfield  {author} {\bibinfo {author} {\bibfnamefont {A.}~\bibnamefont
  {Buchel}}\ and\ \bibinfo {author} {\bibfnamefont {R.~C.}\ \bibnamefont
  {Myers}},\ }\href {\doibase 10.1088/1126-6708/2009/08/016} {\bibfield
  {journal} {\bibinfo  {journal} {JHEP}\ }\textbf {\bibinfo {volume} {08}},\
  \bibinfo {pages} {016} (\bibinfo {year} {2009})},\ \Eprint
  {http://arxiv.org/abs/0906.2922} {arXiv:0906.2922 [hep-th]} \BibitemShut
  {NoStop}%
\bibitem [{\citenamefont {Myers}\ \emph {et~al.}(2007)\citenamefont {Myers},
  \citenamefont {Starinets},\ and\ \citenamefont {Thomson}}]{Myers:2007we}%
  \BibitemOpen
  \bibfield  {author} {\bibinfo {author} {\bibfnamefont {R.~C.}\ \bibnamefont
  {Myers}}, \bibinfo {author} {\bibfnamefont {A.~O.}\ \bibnamefont
  {Starinets}}, \ and\ \bibinfo {author} {\bibfnamefont {R.~M.}\ \bibnamefont
  {Thomson}},\ }\href {\doibase 10.1088/1126-6708/2007/11/091} {\bibfield
  {journal} {\bibinfo  {journal} {JHEP}\ }\textbf {\bibinfo {volume} {11}},\
  \bibinfo {pages} {091} (\bibinfo {year} {2007})},\ \Eprint
  {http://arxiv.org/abs/0706.0162} {arXiv:0706.0162 [hep-th]} \BibitemShut
  {NoStop}%
\end{thebibliography}%

\end{document}